\documentclass[usegraphicx,usenatbib]{mn2e}
\pdfoutput=1
\usepackage[totalwidth=480pt,totalheight=680pt]{geometry}
\usepackage{amssymb}
\usepackage{color}

\newcommand{\lya}{\ifmmode\mathrm{Ly}\alpha\else{}Ly$\alpha$\fi}
\newcommand{\igm}{\ifmmode\mathrm{IGM}\else{}IGM\fi}
\newcommand{\bao}{\ifmmode\mathrm{BAO}\else{}BAO\fi}
\newcommand{\hi}{\ifmmode\mathrm{HI}\else{}HI\fi}
\newcommand{\hii}{\ifmmode\mathrm{HII}\else{}HII\fi}
\newcommand{\hei}{\ifmmode\mathrm{HeI}\else{}HeI\fi}
\newcommand{\heii}{\ifmmode\mathrm{HeII}\else{}HeII\fi}
\newcommand{\cmb}{\ifmmode\mathrm{CMB}\else{}CMB\fi}
\newcommand{\qso}{\ifmmode\mathrm{QSO}\else{}QSO\fi}

\title[Large volume \lya{} forest simulations]
{Fast, large volume, GPU enabled simulations for the $\bmath{\lya{}}$ forest:
  power spectrum forecasts for baryon acoustic oscillation
  experiments}

\author[B. Greig et al.] {Bradley~Greig,$^{1,2}$\thanks{E-mail:~bgreig@student.unimelb.edu.au~(BG),}, James S.~Bolton,$^{1,2}$\thanks{~jsbolton@unimelb.edu.au~(JSB)} \& J. Stuart B.~Wyithe;$^{1,2}$\thanks{~swyithe@unimelb.edu.au~(JSBW),} \\ $^1$School of Physics, University of Melbourne, Parkville, Victoria 3010, Australia\\
$^2$ARC Centre of Excellence for All-sky Astrophysics (CAASTRO)}

\begin{document}
\maketitle \begin{abstract}

\noindent
High redshift measurements of the baryonic acoustic oscillation scale
(\bao{}) from large \lya{} forest surveys represent the next frontier
of dark energy studies. As part of this effort, efficient simulations
of the \bao{} signature from the \lya{} forest will be required. We
construct a model for producing fast, large volume simulations of the
\lya{} forest for this purpose. Utilising a calibrated semi-analytic
approach, we are able to run very large simulations in 1 Gpc$^3$
volumes which fully resolve the Jeans scale in less than a day on a
desktop PC using a GPU enabled version of our code. The \lya{} forest
spectra extracted from our semi-analytical simulations are in
excellent agreement with those obtained from a fully hydrodynamical
reference simulation. Furthermore, we find our simulated data are in
broad agreement with observational measurements of the flux
probability distribution and 1D flux power spectrum.  We are able to
correctly recover the input \bao{} scale from the 3D \lya{} flux power
spectrum measured from our simulated data, and estimate that a
BOSS-like $10^{4}\rm~deg^{2}$ survey with $\sim 15$ background sources
per square degree and a signal-to-noise of $\sim 5$ per pixel should
achieve a measurement of the \bao{} scale to within $\sim$1.4 per
cent.  We also use our simulations to provide simple power-law
expressions for estimating the fractional error on the \bao{} scale on
varying the signal-to-noise and the number density of background
sources. The speed and flexibility of our approach is well suited for
exploring parameter space and the impact of observational and
astrophysical systematics on the recovery of the \bao{} signature from
forthcoming large scale spectroscopic surveys.

\end{abstract} 

\section[Introduction]{Introduction}

The \lya{} forest is the series of absorption lines arising from the
resonant scattering of redshifted \lya{} photons emitted from a
background source by the intervening neutral hydrogen in the
intergalactic medium \citep{Rauch:1998p4563}. The neutral hydrogen is
in photo-ionisation equilibrium with the ionising background produced
by the integrated emission from galaxies and quasars \citep[e.g.][]{Meiksin:2009p6791}. Numerical simulations and
semi-analytical models have demonstrated that the \lya{} forest is a
valuable probe of the underlying matter density field in the
intergalactic medium (IGM), tracing the so-called ``cosmic web'' of
large scale structure \citep{Cen:1994p7419,Hernquist:1996p4477,Bi:1997p3607,Theuns:1998p7245}. The measurement of fluctuations in the
transmitted \lya{} flux in samples of quasar spectra
\citep{McDonald:2000p388,Croft:2002p6751,Kim:2004p7039}, combined with a suitable model to connect
the observed flux to the underlying matter distribution, thus enable
the matter power spectrum (PS) to be inferred on small scales along
the line-of-sight \citep{Croft:2002p6751,Viel:2004p7045,McDonald:2005p408}.

In recent years, it has been proposed that the \lya{} forest of
absorption lines may also be used to detect the signature of baryonic
acoustic oscillations (\bao{}) in the large scale structure of the
\igm{} \citep[e.g.][]{McDonald:2007p6752}. BAOs provide a standard
ruler which may be used to examine the geometry of the Universe and
the nature of dark energy, and are already observed in the Cosmic
Microwave Background (\cmb{}) at $z\sim1100$ and the clustering of
galaxies at low redshift \citep{Eisenstein:2005p5087,Cole:2005p5206,Hutsi:2006p6079,Blake:2007p6108,Padmanabhan:2007p6159,Percival:2007p5923,Percival:2010p6176,Beutler:2011p10735,Blake:2011p10710}. The
characteristic \bao{} signal is spatially correlated on scales of
order $\sim150$ comoving Mpc; large survey volumes are therefore required to
provide adequate statistics for the detection of this scale. Existing
low redshift studies are subject to a degeneracy between the
space-time curvature $\Omega_{k}$ and an evolving dark energy equation
of state \citep{2005AAS...207.2605Z}.  Studying \bao{} at higher redshift
can help alleviate this difficulty.  However, the extension of galaxy
surveys to higher redshifts becomes increasingly expensive because of
the significant telescope resources required to observe a sufficient
number of (fainter) galaxies.

An alternate measurement of large scale structure at $z \simeq 2 - 3$
is provided by the \lya{} forest. There are several advantages to
using the \lya{} forest as a probe of \bao{}. At higher redshifts the
evolution of $\Omega_{k}$ and dark energy differ ($\Omega_{k}$ evolves
as $\propto(1+z)^{2}$, whereas dark energy is expected to evolve more
slowly with redshift) allowing one to break this degeneracy and to
obtain constraints on the dark energy equation of state at this
epoch. Furthermore, the detection of suitable background sources
becomes significantly easier due to the peak in quasar number density
observed at $z\simeq2.2$
\citep{Richards:2006p8099}. \citet{McDonald:2007p6752} demonstrated
that low to medium signal-to-noise (S/N) spectra of a large number of
quasars with sufficient density per square degree could be used to
detect the \bao{} signal with the \lya{} forest. As an example, these
authors estimated that the \bao{} scale could be measured in a 2000
square degree survey with 40 quasars per square degree. The \bao{}
signal may then be statistically extracted from such a sample via
either the correlation function (measured as a characteristic peak) or
the power spectrum (measured as a characteristic oscillation
period). More recently \citet{McQuinn:2011p10709} demonstrated the
sensitivity of future \lya{} forest surveys to the flux correlation
function, and investigated the optimal survey configuration for
estimating the \lya{} forest correlations including the use of
galaxies as background sources to boost the \lya{} forest survey
sensitivity. These authors also estimate the sensitivity of a future
\bao{} measurement to the systematics associated with \lya{} forest
surveys.

In anticipation of the necessary large volume \lya{} forest surveys
such as the Baryon Oscillation Spectroscopic Survey \citep[BOSS;][]{Schlegel:2009p6826,Slosar:2011p8013}, simulation work has also been carried
out to construct synthetic \lya{} forest spectra for use in
constructing mock surveys. Recently, large volume, high resolution
dark matter (DM) N-body simulations have been performed by
\citet{White:2010p5647} and \citet{Slosar:2009p5710}, while
\citet{Norman:2009p5576} have presented fully hydrodynamical
simulations in slightly smaller volumes for this purpose. In this work
we outline a method for producing fast (\textit{i.e.} less
than one day), similarly large volume, high resolution simulations using a
single desktop PC with a graphics processing unit (GPU). Our approach,
which is based on widely used semi-analytical models for the \lya{}
forest \citep{Bi:1993p69,Reisenegger:1995p7082,Gnedin:1996p295,Bi:1997p3607,Hui:1997p4783,Gnedin:1998p83,Choudhury:2001p4404,Matarrese:2002p3671,Viel:2002p35,Viel:2002p4288}, does not capture
the mildly non-linear effects of the \lya{} forest modelled in the
N-body and hydrodynamical simulations. Nevertheless, this method is
well suited for studying the statistics of \lya{} forest absorption on
large scales, where the assumption of linear evolution is a reasonable
approximation. Our approach is therefore complimentary to more
accurate but very expensive numerical simulations. We note another
semi-analytical approach for studying the \bao{} in the \lya{} forest
has also been recently presented by \citet{Kitaura:2010p7306}.

This paper is organised as follows. In Section \ref{Sec:Model}, we provide a summary
of the semi-analytical model and the method used for generating our
synthetic \lya{} forest spectra. In Section \ref{Sec:Performance} we compare the semi-analytical density and velocity fields to a fully hydrodynamical simulation. In Section \ref{Sec:Sims} we describe the
simulations used in this work in more detail. In Section \ref{Sec:Comparison} we test our
model by comparison to a fully hydrodynamical simulation and a
selection of observational data, and in Section \ref{Sec:Recovery} we extract the
\bao{} signature from a mock \lya{} \bao{} survey and compare our
approach to other recently published work. In Section \ref{Sec:Stats} we provide scaling relations for the fractional error on the recovery of the \bao{} scale for \lya{} forest \bao{} surveys. Finally, in Section \ref{Sec:Conclusion} we
finish with our closing remarks. An appendix detailing the performance
and implementation of our GPU enabled code is included at the end of
the paper.

\section[Semi-Analytical Model for the IGM]{Semi-Analytical Model for the \igm{}} \label{Sec:Model}

Many approximate techniques have been proposed for modelling the
mildly non-linear, low column density \igm{} from an initial dark
matter distribution. Approaches taken include the lognormal method
\citep{Coles:1991p4666} applied to the linear DM distribution to
mimic non-linear behaviour \citep{Bi:1993p69,Bi:1997p3607,Choudhury:2001p4404}, the
rank-ordered mapping of linear to non-linear densities using a
calibration hydrodynamical simulation \citep{Viel:2002p4288} or
using the Zel'dovich approximation \citep{ZelDovich:1970p7102} to
generate the DM distribution. Many of these models also subsequently
smooth the initial DM density field on a scale related to the Jeans
length in the low density \igm{}, accounting for the effect of gas
pressure on the baryon distribution on small scales
\citep{Reisenegger:1995p7082,Gnedin:1996p295,Gnedin:1998p83,Hui:1997p4783,Matarrese:2002p3671,Viel:2002p35}.

In this work we follow \citet{Viel:2002p4288}, who investigated
various approaches for producing more accurate models for the gas
density distribution using semi-analytical models. These authors found
they could better mimic the non-linear DM distribution by taking the
DM density probability distribution function (PDF) from a linear
simulation and performing a rank-ordered mapping to the corresponding
distribution obtained from a hydrodynamical simulation.  We adopt a
similar approach here, and use a hydrodynamical simulation to
calibrate the one point distribution for the density field. This
ensures we obtain accurate one and two point statistics for most of
the absorption in the resulting \lya{} spectra. However, as noted by
\citet{Viel:2002p4288} the two point statistics for regions of strong
absorption with $F = e^{-\tau}<0.1$, where non-linear effects are
important, will not be properly captured by this model. In this paper,
we demonstrate this limitation will not present a serious impediment
to extracting the \bao{} signal on large scales from our simulated
data.  We stress, however, that our semi-analytical simulations are
based on linear theory and they do not include large scale non-linear
evolution and \bao{} damping as described in the detailed N-body work
of \cite{Seo:2007p7988}.  Rather, our models can be used to address
the non-gravitational issues associated with the \lya{} forest, and
will need to be coupled with N-body studies of the gravitational
evolution of the \bao{} in order to make detailed comparisons with
real data.

We use model L3 of \citet{Bolton:2010p6750} as our calibration
hydrodynamical simulation, with a snapshot at $z=2.976$ generated
using the parallel Tree-SPH code {\sc{GADGET-3}}
\citep{Springel:2005p7117}. This simulation assumes a $\Lambda$CDM
cosmology with, $h=0.72$, $\Omega_{\rm{m}} = 0.26$,
$\Omega_{\Lambda}=0.74$, $\Omega_{\rm{b}}=0.0444$, $n_{\rm{s}}=0.96$
and $\sigma_{8}=0.8$. The calibration simulation has a box size of 40
$h^{-1}$ Mpc and contains $2\times512^{3}$ gas and DM particles,
yielding a gas particle mass resolution of $5.9\times10^{6} h^{-1}
M_{\odot}$. Importantly, this particular box size and mass resolution
is sufficient for resolving the gas densities responsible for the
\lya{} forest at $z\simeq3$ \citep{Bolton:2009p4777}. We note
  that in this work we assume a constant redshift of $z=3$ for our
  simulations. The accuracy to which the \bao{} scale can be recovered
  is redshift dependent and as such we would expect our results to
  differ slightly across the redshift range of $z=2-3$.

\subsection[Generation of the density field]{Generation of the density field}

We begin by generating a linear DM density field in Fourier space
within a cubic simulation volume according to the transfer function of
\citet{Eisenstein:1998p3096}. The density field is then linearly
evolved using the linear growth factor $D_{+}(z)$ to the redshift of
interest. Following \citet{Bi:1997p3607}, we account for the
effect of gas pressure on the baryons by smoothing the linear DM
density field with the kernel

\begin{equation}
\delta_{\rm{b}}(\bmath{k},z) = \frac{\delta_{\rm{DM}}(\bmath{k},z)}{1+k^{2}/k^{2}_{F}(z)},
\end{equation}

\noindent
where $k_{F}$ is the filtering scale, which is related to the comoving
Jeans scale $k_{J}$ via $k_{F} = \epsilon k_{J}$. The comoving Jeans
scale is given by

\begin{equation}
k_{J} = H_{0}\left[\frac{3\mu m_{\rm{p}}\Omega_{\rm{m}}(1+z)}{2\gamma k_{\rm{B}}T(z)}\right]^{1/2},
\end{equation}

\noindent
where $\mu$ is the reduced mass and $T(z)$ is the gas temperature. The
smoothing scale $k_F$ is a free parameter in the simulations, and
effectively accounts for the finite delay between heating and the
subsequent pressure response of the gas, which is dependent on the
specific reionisation history
\citep{Gnedin:1998p83,Desjacques:2005p3507}. In this work we choose
$k_{F} = 6.5$ Mpc$^{-1}$, which we find provides good
agreement\footnote{\citet{Viel:2002p4288} account for gas pressure by
  using a third order polynomial fit to the relationship between the
  baryon and DM density in their calibration hydrodynamical
  simulation, rather than applying a global smoothing to the density
  field in Fourier space as we do. However, we found that applying a
  global smoothing scale to the linear DM density field, and then
  mapping from the DM density PDF directly to the baryon PDF of the
  hydrodynamical simulation, produced better agreement for both
  individual lines of sight and for \lya{} flux statistics.} with
spectra extracted from our calibration hydrodynamical simulation and
the observational data (see Section \ref{Sec:Comparison}). In
  Section \ref{Sec:Performance} we compare the generated linear
  density field and the corresponding non-linear rank-ordered density
  field to the calibration hydrodynamical simulation using the three
  dimensional matter power spectrum.

\subsection[Generation of the velocity field]{Generation of the velocity field}

In addition to the density field, the \lya{} forest is also sensitive
to the peculiar velocity field. We generate the linear peculiar
velocity field, based on our linear DM density field, using the
following expression,
\begin{equation} \label{eq:velocity}
\bmath{v}_{IGM}(\bmath{k},z) = E_{+}(z)i\bmath{k}/k^{2}\delta_{\rm{DM}}(\bmath{k}),
\end{equation}
where $\bmath{v}_{IGM}$ is expressed relative to the comoving
wavenumber $\bmath{k}$, $E_{+}(z) =
H(z)f(\Omega_{\rm{m}},\Omega_{\rm{\Lambda}})D_{+}(z)/(1+z)$ and
$f(\Omega_{\rm{m}},\Omega_{\rm{\Lambda}}) = -\rm{dln}
D_{+}(z)/\rm{dln} (1+z)$. We compare the linear peculiar
  velocities generated from Equation \ref{eq:velocity} with the
  peculiar velocities in the calibration hydrodynamical simulation in
  more detail Section \ref{Sec:Performance}.

\subsection[Generation of Ly-alpha forest spectra]{Generation of $\bmath{\lya{}}$ forest spectra}

The calibrated baryonic density and peculiar velocity fields are next
used to generate the synthetic \lya{} forest spectra. We use the
approach outlined in \citet{Hui:1997p4783}, assuming that the neutral
hydrogen in the ionised \igm{} is in photo-ionisation equilibrium;
this should be a reasonable approximation at $z\simeq3$. The proper
number density of neutral hydrogen in the \igm{} is

\begin{eqnarray}
n_{\rm{HI}}(\bmath{x},z) & = & 7.24\times10^{-6}\bar{n}_{\rm{H}}(z)\left(\frac{T}{10^{4}K}\right)^{-0.7}\left(\frac{\Omega_{\rm{b}}h^2}{0.024}\right) \nonumber \\
 & & \times \left(\frac{1}{\Gamma_{-12}}\right)[1+\delta_{\rm{b}}(\bmath{x},z)]^{2}\left(\frac{1+z}{4}\right)^{3},
\end{eqnarray}

\noindent
where $\bar{n}_{\rm{H}}$ is the proper mean density of neutral
hydrogen, $\Gamma_{-12}$ is the hydrogen photo-ionisation rate in units
of $10^{-12}\rm\,s^{-1}$ and $\delta_{\rm{b}}$ is the baryonic
overdensity and is a function of the comoving position $\bmath{x}$. We
take the case-A recombination rate of neutral hydrogen to be
$4.17\times10^{-13} \left(\frac{T}{10^4
  K}\right)^{-0.7}\rm\,cm{^3}\,s^{-1}$. The temperature of the low
density \igm{}, $\delta + 1 \leq 10$, is then modelled assuming a
power-law temperature-density relation

\begin{equation} \label{eq:tempdens}
T = T_{0}(1+\delta_{\rm{b}})^{\gamma-1},
\end{equation}
where $\gamma$ is the polytropic index describing the slope of the
temperature-density relation. This relation is expected to arise
following reionisation due to the interplay between photo-heating and
adiabatic cooling, where typically $1 < \gamma < 1.6$
\citep{Hui:1997p4783,Valageas:2002p7220}. However, at densities
$\delta + 1 > 10$, radiative cooling becomes more efficient and the
power-law temperature density relation is no longer a good
approximation. We therefore employ a pivot point at $\delta + 1 = 10$,
below which the typical temperature-density relation still holds, but
above which we set the temperature to be constant, such that $T(\delta
+ 1>10)=T(\delta + 1= 10)$. Note, however, shock heating will
introduce some scatter into this relation. Furthermore, \heii{}
reionisation, which is expected to end around $z\simeq 3$, may also
produce a relationship between temperature and density which is more
complicated than this tight power-law
\citep{Bolton:2009p7229,McQuinn:2009p7230}. There is also some
observational evidence for $\gamma<1$ ({\it i.e.} an inverted
temperature-density relation, \citealt{Viel:2009p7711}), although it
appears difficult to achieve this via heating during \heii{}
reionisation alone
\citep{Bolton:2009p7229,McQuinn:2009p7230}. However, we defer the
discussion of such possible systematic uncertainties to a future
study.

We next generate the transmitted \lya{} flux along the line-of-sight
for our synthetic \lya{} forest spectra using the relation $F =
e^{-\tau}$, where $\tau$ is the \lya{} optical depth. The optical
depth of the synthetic \lya{} spectra is computed using
\citep[e.g.][]{Theuns:1998p7245}
\begin{equation}
\tau_{\rm{\alpha}}(i) =\frac{c\sigma_{\alpha}\delta R}{\pi^{1/2}}\sum_{j=1}^{N} \frac{n_{\rm{HI}}}{b_{\rm{HI}}(j)} \exp\left[-\left(\frac{v_{\rm{H}}(i)-u(j)}{b_{\rm{HI}}(j)}\right)^{2}\right],
\end{equation}
where $i$ and $j$ denote pixels along the line of sight through the
simulation volume, $\delta R$ is the pixel width in proper
coordinates, $\sigma_{\alpha} = 4.48\times10^{-18}\, \rm{cm}^{2}$ is
the scattering cross-section for $\lya$ photons, $b_{\rm{HI}} =
\left(\frac{2k_{\rm{B}}T}{m_{\rm{H}}}\right)^{1/2}$ is the Doppler
parameter describing the thermal width of the line profiles, $v_{H}$
is the Hubble velocity and $u(j)$ is the total velocity given by the
summation of the Hubble flow and the peculiar velocity along the line
of sight, $u(j) = v_{\rm{H}}(j)+v_{\rm{pec}}(j)$.

Finally, once we have our optical depth along the line of sight, we
renormalise the optical depths of all of our spectra to match the
observed mean flux of the \lya{} forest at $z\simeq3$. We define the
transmitted flux as $\langle{F}\rangle = \langle e^{-A\tau}\rangle$,
where $A$ is a normalisation constant to be solved for. This
modification is equivalent to rescaling the \hi{} photo-ionisation
rate produced by the UV background. We solve for $A$ by summation over
all generated spectra and iterate $A$ until the mean transmitted flux
from the simulated spectra matches the observationally measured
value. We match our mean transmitted flux to that observed by
\citet{Kim:2007p3619}, corresponding to a mean transmitted flux of
$\langle{F}\rangle = 0.72$ or an effective optical depth of
$\tau_{\rm{eff}} = 0.329$ at $z=3$.  Lastly, we note that in the
generation of our \lya{} spectra, we do not include the redshift evolution
of the effective optical depth along individual lines-of-sight.  The
effect of metal absorption lines and the damping wings originating
from high column density absorption systems are also excluded.

\section[Semi-Analytic model performance]{Semi-Analytic model performance} \label{Sec:Performance}

Before comparing the \lya{} forest simulations used in this
  work with observations, we first verify the performance of our
  rank-ordered semi-analytic model. Firstly we investigate the effect
  that rank-ordered mapping of the linear density field has on the
  matter power spectrum, and secondly, we check that our linear
  peculiar velocity field produces a reasonable description of the
  peculiar velocity field when compared to hydrodynamical simulations.

\begin{figure} 
	\begin{center}
		\includegraphics[trim = 2.5cm 1.5cm 2.2cm 2cm, scale = 0.32]{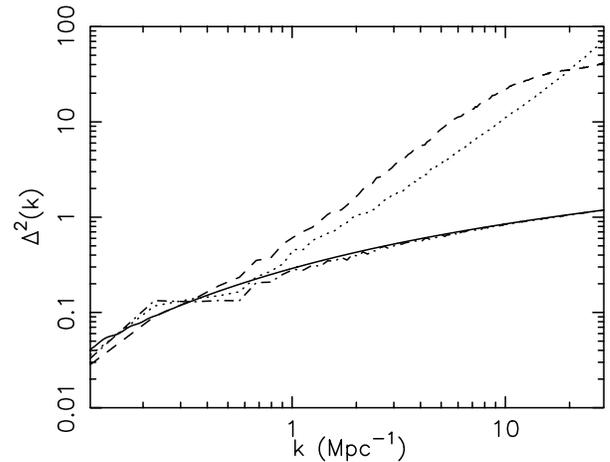}
	\end{center}
\caption{The dimensionless dark matter PS of our 40 $h^{-1}$ Mpc, 512$^{3}$ semi-analytical simulation before (dot-dashed) and after (dotted) the rank-ordered mapping of the linear density field with our calibration hydrodynamical simulation. For comparison we also show the theoretical linear input PS (solid) and the dark matter PS of the calibration hydrodynamical simulation (dashed).}
\label{fig:DensityPS}
\end{figure}

\subsection[Matter power spectrum]{Matter power spectrum}

In Figure \ref{fig:DensityPS}, we illustrate the effect of the
  rank-ordering procedure on the linear dark matter density field by
  comparing the dimensionless dark matter PS from our semi-analytic
  simulations to the dark matter PS from the calibration
  hydrodynamical simulation. We find that the rank-ordered linear
  matter power spectrum correctly recovers the large scale behaviour
  when compared to the non-linear dark matter PS from the
  hydrodynamical simulations. For smaller spatial scales, the
  non-linear density from the rank-ordered method underpredicts the
  correct non-linear behaviour. However underproducing the small-scale
  power will have no significant effect on the \bao{} scale. Thus,
  provided we are able to maintain roughly the same spatial resolution
  in our large-scale simulations as in our calibration simulation, the
  rank-ordered mapping procedure will perform well at reproducing the
  correct large-scale behaviour required for accurately simulating the
  recovery of the \bao{} scale.

\begin{figure*} 
	\begin{center}
		\includegraphics[trim = 1.5cm 11cm 2.2cm 1.5cm, scale = 0.65]{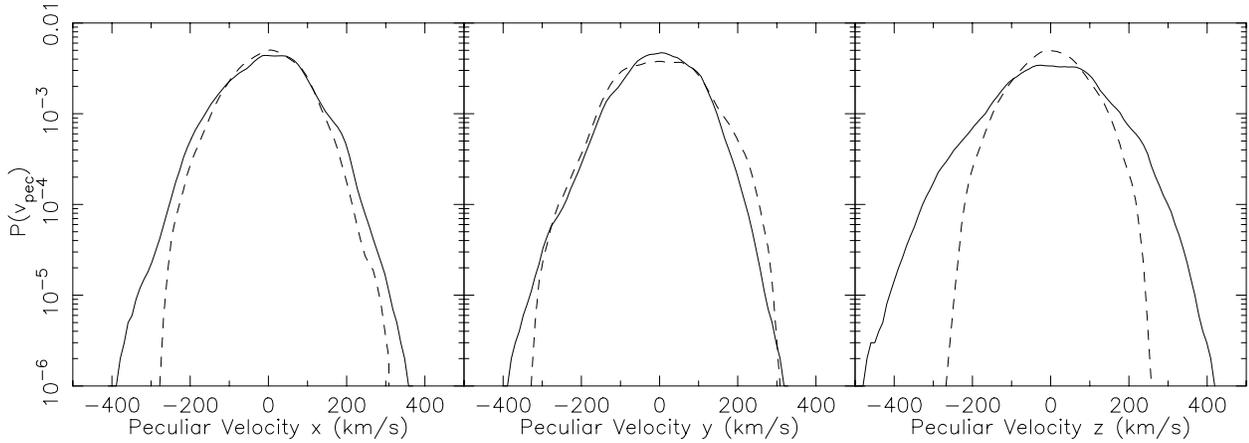}
	\end{center}
\caption{The linear peculiar velocity PDF's of our 40 $h^{-1}$
    Mpc, 512$^{3}$ semi-analytical simulation. From left to right are
    the peculiar velocity PDF's in the x, y and z directions of our
    simulation (dashed) and the calibration hydrodynamical simulation
    (solid). The linear peculiar velocity field matches reasonably
    well in the low density regions which have smaller peculiar
    velocities.  However, the semi-analytical model does not produce
    the larger (rarer) peculiar velocities expected in non-linear,
    overdense regions.}
\label{fig:VelocityPDFs}
\end{figure*}


\subsection[Peculiar velocities]{Peculiar velocities}
As discussed in the previous section, we also produce the
  linear peculiar velocity field prior to the rank-ordered mapping of
  the density field.  In Figure \ref{fig:VelocityPDFs} we compare the
  resulting linear peculiar velocity PDFs to those from the
  hydrodynamical simulation. We find the linear peculiar velocity
  field to be isotropic, and that the velocities from our simulations
  match quite well in the low density regions. However as expected,
  higher density regions with larger (but rarer) peculiar velocities
  (which correspond to regions of infall in the hydrodynamical
  simulation) are not correctly captured in this model.

\begin{figure*} 
	\begin{center}
		\includegraphics[trim = 2cm 1cm 2.2cm 2cm, scale = 0.66]{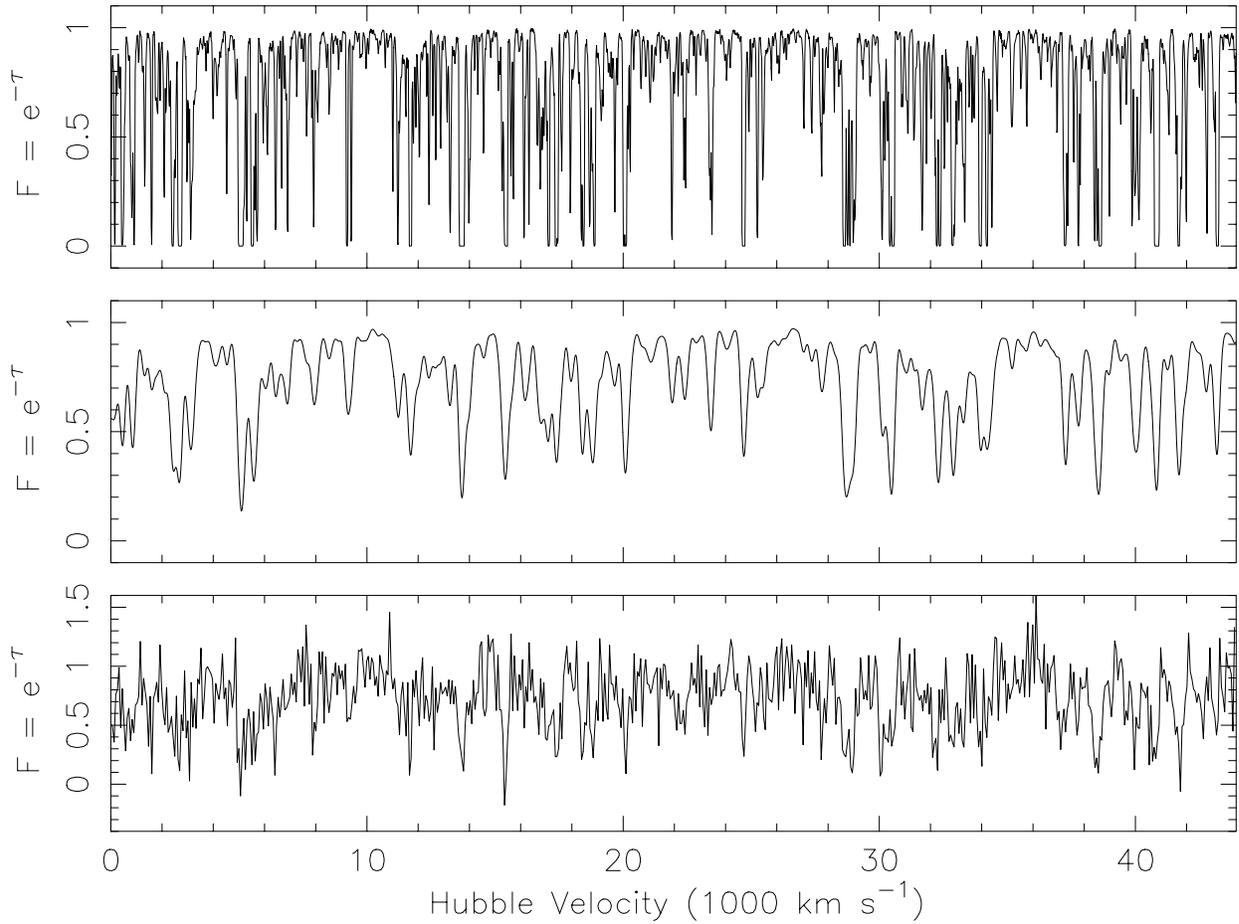}
	\end{center}
\caption{Example synthetic \lya{} forest spectrum generated from our
  $1\rm\,Gpc^{3}$ simulation with 4096$^3$ pixels. {\it Upper panel:}
  The transmitted flux along the line-of-sight predicted by the
  simulation. {\it Middle panel:} The same spectrum after convolution
  with a Gaussian instrument profile with
  FWHM$=224\rm\,km\,s^{-1}$. {\it Lower panel:} The fully processed
  spectrum, resampled onto pixels of width $\sim 1\,$\AA~with Gaussian
  distributed noise corresponding to a signal-to-noise of $5$ per
  pixel.  This spectrum is representative of the mock data set we use
  to recover the \bao{} signal from the \lya{} forest. Note that the
  redshift evolution of the effective optical depth along the
  line-of-sight is not included in this model.}
\label{fig:LowResQSOConvolved}
\end{figure*}

\section[Ly-alpha forest simulations]{$\bmath{\lya{}}$ forest simulations} \label{Sec:Sims}

We construct two different \lya{} forest simulations in this work; a
high resolution simulation for comparison to our reference
hydrodynamical simulation, and a low resolution, large volume
simulation for recovery of the \bao{} signature. Both
models are compared to published measurements of the \lya{} flux
probability distribution function (PDF) and the 1D flux power spectrum
(PS). In both simulations we assume a temperature-density relation
with $\gamma = 1.3$ and we set the temperature $T_{0} =
1.7\times10^4~$K, broadly consistent with the observational constraints
on the \igm{} thermal state at $z=3$ \citep{Schaye:2000p7569,Lidz:2010p7574,Becker:2011p7578}.

\subsection[High resolution simulations]{High resolution simulations} \label{subsec:HighRes}

In our high resolution model, we simulate a 40 $h^{-1}$ Mpc simulation
box, containing $512^{3}$ pixels. The box size and resolution are
chosen to mimic our calibration {\sc{GADGET-3}} hydrodynamical
simulation, although we note that the resolution comparison will not
be exact due to the spatially adaptive resolution of
{\sc{GADGET-3}}. In order to compare our simulated \lya{} forest
spectra to observed high resolution data, we must also process our
simulated spectra to mimic the properties of the data. We convolve the
spectra with a Gaussian with a FWHM $=7\rm~km~s^{-1}$, and resample
our spectra onto 0.05~\AA~bins. We finally add Gaussian distributed
noise assuming S/N $\sim$ 50 per pixel. The same procedure is also
performed on the \lya{} spectra generated from the hydrodynamical
simulation.

\subsection[Low resolution, large volume simulations]{Low resolution, large volume simulations}

Recovery of the \bao{} signal from our simulation requires that we
also simulate two large volume simulations at lower resolution. One
simulation is generated using a matter PS containing baryon
oscillations, and the other has the baryon oscillations suppressed. We
use the transfer functions of \citet{Eisenstein:1998p3096} for this
purpose. Each simulation is generated in a $1\rm\,Gpc^{3}$ comoving
box, containing 4096$^3$ pixels \citep[chosen to be comparable with the 4000
pixels per spectra computed using the roadrunner supercomputer by][]{White:2010p5647}. Each pixel is therefore $\sim244$ comoving
kpc. In comparison, the comoving Jeans smoothing scale (given by
equation 2) at $z=3$ is $\sim760\rm\,kpc$ (and our filtering scale
$k_{F} = 6.5 ~\rm{Mpc}^{-1}$ corresponds to a comoving scale of $\sim
966\rm\,kpc$). Importantly, this implies our large volume simulations
adequately resolve the Jeans scale at mean density.

In order to mimic the low resolution \lya{} forest data expected in
forthcoming \bao{} surveys we also convolve our spectra with a
Gaussian with a FWHM of $\sim3.63$~\AA~( $\sim 224\rm\,km\,s^{-1}$) and
resample the spectra onto 1.0375~\AA~bins. These values are
representative of BOSS spectra \citep{Eisenstein:2011p7600}. We
also add Gaussian distributed noise, ${\rm S/N}=5$ per pixel. Finally
we ensure each synthetic line of sight corresponds to the pathlength
between the quasar rest frame \lya{} and Ly$\beta$ transitions only,
minus the $3000$ km s$^{-1}$ blueward of \lya{} in the quasar rest
frame. The latter accounts for the quasar proximity effect in the
observational data. An example \lya{} forest sightline drawn from our
simulation is displayed in Figure \ref{fig:LowResQSOConvolved}. It is
these spectra which will be used in the \bao{} recovery described in
Section \ref{Sec:Recovery}.  Before this, however, we now proceed to perform
consistency checks on our simulation output by comparing the synthetic
\lya{} forest spectra to measurements of \lya{} flux statistics.

\section[Flux Statistics]{Flux Statistics} \label{Sec:Comparison}
\subsection[Available data]{Available data}

We shall compare our simulations to the measured flux PDF
\citep{McDonald:2000p388,Kim:2007p3619,Desjacques:2007p7310} and the 1D line-of-sight flux PS
\citep{McDonald:2000p388,Croft:2002p6751}. The
\citet{Kim:2007p3619} sample contains 18 high resolution quasar
spectra obtained with the VLT/Ultraviolet and Visual Echelle
Spectrograph (UVES), specifically chosen to have a signal-to-noise of
at least $30-50$ and to fully sample the \lya{} forest region.  In
comparison, the \cite{McDonald:2000p388} sample contains 8 high
resolution quasar spectra obtained using the High Resolution Echelle
Spectrometer (HIRES) at the Keck telescope. However, the
\citet{Croft:2002p6751} sample contains both high resolution spectra
from Keck HIRES and 23 low resolution spectra obtained with the Low
Resolution Imaging Spectrometer (LRIS). The
\citet{Desjacques:2007p7310} sample contains 3492 low resolution
quasar spectra from the Sloan Digital Sky Survey (SDSS) data release
three (DR3).

\subsection[The flux probability distribution function]{The flux probability distribution function}

\begin{figure} 
	\begin{center}
		\includegraphics[trim = 1cm 2cm 0.5cm 3cm, scale = 0.32]{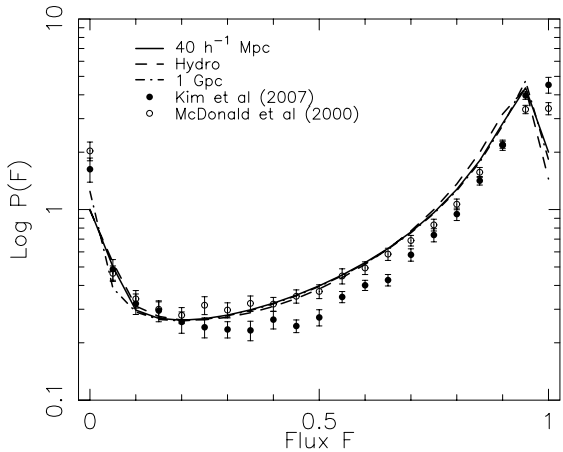}
	\end{center}
\caption{The flux PDF generated from our 40 $h^{-1}$ Mpc, 512$^3$
  semi-analytical simulation (solid curve) and our calibration
  hydrodynamical simulation (dashed curve), compared to observational
  measurements made using high resolution quasar spectra by
  \citet{Kim:2007p3619} (closed circles) and \citet{McDonald:2000p388}
  (open circles) at $z\simeq 3$. The simulated data has been processed
  to resemble the resolution and S/N of the observational data. For
  comparison, the flux PDF from our large volume (1 Gpc, 4096$^3$)
  semi-analytical simulation is also displayed (dot-dashed curve). The
  spectra have not been processed to resemble observational data in
  this latter instance.}
\label{fig:FluxPDF}
\end{figure}

We first compare the flux PDF constructed from our synthetic \lya{}
forest spectra to measurements at $z\simeq 3$ obtained from high
resolution data in Figure \ref{fig:FluxPDF}. The data points
correspond to the measurements presented by \citet{Kim:2007p3619} and
\citet{McDonald:2000p388}. Note that \citet{Kim:2007p3619} and
\citet{McDonald:2000p388} use different prescriptions for the removal
of the metal lines. The flux PDF measurement performed by
\citet{Kim:2007p3619} removes suspected metal lines by Voigt profile
fitting, whereas the \citet{McDonald:2000p388} sample instead excises
regions which are suspected of being contaminated by metal lines.

Following the observational measurements, we compute the flux PDF from
our simulations by separating the transmitted flux into $21$ equally
spaced flux bins of width $\Delta F = 0.05$, from $F = 0$ to $F=1$
(i.e. at $F = 0$, the first data point contains flux from $-0.025 \leq
F < 0.025$). The solid and dashed curves in Figure \ref{fig:FluxPDF}
correspond to the flux PDF computed from the synthetic \lya{} spectra
extracted from our high resolution semi-analytical simulation and the
full hydrodynamical simulation, respectively. These spectra have been
processed to resemble the observational data, as described in Section
\ref{subsec:HighRes}. The dot-dashed curve instead shows the flux PDF computed from the
spectra extracted from the $1\rm~Gpc^{3}$ simulation box before the data
are processed to resemble the low resolution data.

The PDF generated by our high resolution semi-analytical simulation
matches remarkably well with the hydrodynamical simulation, with only
a small difference observed in the PDF at $F=0.8$--$1$. Furthermore,
the simulations are also in broad agreement with the
\citet{McDonald:2000p388} data, although they do not agree so well
with the more recent observations of \citet{Kim:2007p3619}. On the
other hand it has been shown by \citet{Bolton:2008p5404} that a better
match to the observational data of \citet{Kim:2007p3619} may be
achieved when an inverted temperature-density relation ($\gamma < 1$)
is assumed; we have instead adopted $\gamma=1.3$ in this work. The
lower resolution, large volume simulation also matches the high
resolution simulations reasonably well.

\begin{figure} 
	\begin{center}
		\includegraphics[trim = 1cm 2cm 0cm 3cm, scale = 0.32]{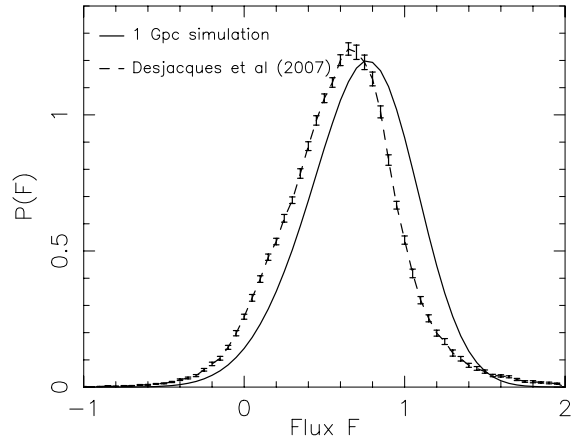}
	\end{center}
\caption{The \lya{} forest flux PDF measured from the low resolution
  SDSS DR3 data at $z=3$ \citep{Desjacques:2007p7310} is shown by
  the dashed curve with data points. The solid curve compares the
  flux PDF generated from our 1 Gpc semi-analytical simulation after
  the spectra have been modified to match the resolution and noise
  properties of the data. Note that the level of disagreement between
  the simulations and observational data is similar to that found by
  \citet{Desjacques:2007p7310}. These authors noted that altering the
  continuum level on the synthetic spectra can significantly improve
  agreement with the data.}
\label{fig:FluxPDFLowRes}
\end{figure}

As an additional consistency check, in Figure \ref{fig:FluxPDFLowRes}
we also compare the flux PDF constructed from our large, low
resolution simulation to the flux PDF measured from low resolution
data by \citet{Desjacques:2007p7310}. In order to compare our
simulated spectra to the \citet{Desjacques:2007p7310} flux PDF, we
process our simulated spectra to mimic the SDSS DR3 data by convolving
the flux by a Gaussian with FWHM of 170 km s$^{-1}$, resampled onto
$\sim1$\AA~bins and adding Gaussian distributed noise with $\rm
S/N=3.8$ per pixel. In contrast to the high resolution PDF, the
agreement between our simulated low resolution spectra and the
observed flux PDF from the low resolution data is rather poor.
However, \citet{Desjacques:2007p7310} found a very similar
disagreement between their simulations and the flux PDF. These authors
attributed this difference to the single power law approximation used
for the quasar continuum level in the observational data.
\citet{Desjacques:2007p7310} found they could improve the agreement
between their observations and simulation by introducing a break in
the continuum slope, with a decrease in the mean quasar continuum
($\sim10-15$ per cent) and introducing residual scatter ($\sim20$ per
cent) into the continuum level.

Finally, we note that the flux PDF is sensitive to the free parameters
which are inputs to our simulation model. Changes to either the
smoothing scale, $k_{\rm F}$, or the slope of the temperature-density
relation, $\gamma$, in particular will alter the shape of the PDF,
although the flux PDF is relatively insensitive to the assumed
temperature at mean density $T_{0}$ for fixed $\tau_{\rm eff}$
\citep{Bolton:2008p5404}. For example, we could match the data of
\citet{Kim:2007p3619} more closely if we allowed the free parameters
in our model such as $\gamma$ or $k_{F}$ to vary.

\subsection[The 1D flux power spectrum]{The 1D flux power spectrum}

\begin{figure}
	\begin{center}
		\includegraphics[trim = 1cm 2cm 0cm 3cm, scale = 0.32]{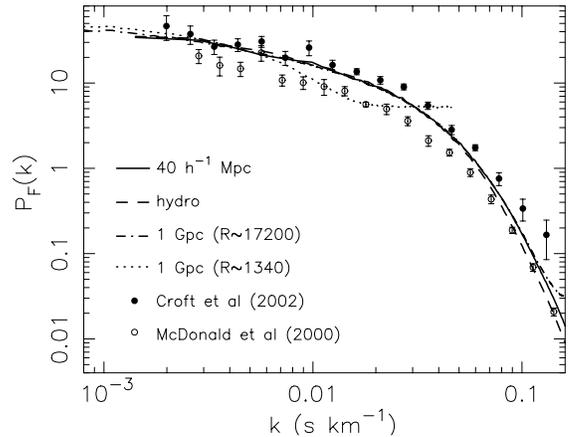}
	\end{center}
\caption{The 1D line-of-sight flux power spectrum generated from our
  high resolution semi-analytical simulation model (solid curve)
  compared to the result from our reference hydrodynamical simulation
  (dashed curve). Observational measurements by
  \citet{McDonald:2000p388} and \citet{Croft:2002p6751} at $z=3$ are
  shown by the open and closed circles, respectively.  The 1D flux PS
  from our large 1 Gpc simulation box before (dot-dashed curve) and
  after (dotted curve) the spectra are degraded to resemble low
  resolution \lya{} forest spectra are also displayed. The degraded
  spectra resemble the line-of-sight displayed in the lower panel of
  Figure \ref{fig:LowResQSOConvolved}.}
\label{fig:FluxPS}
\end{figure}

A more stringent test of the simulated data is the comparison of the
synthetic spectra to higher order flux statistics such as the
line-of-sight flux PS. Figure~\ref{fig:FluxPS} displays the
measurements of the 1D flux PS made by \citet{McDonald:2000p388} and
\citet{Croft:2002p6751} at $z=3$. We generate the 1D flux PS along the
line-of-sight for both the high resolution semi-analytical simulation
(solid curve) and the hydrodynamical simulations (dashed curve) for
comparison to the data. The simulated results are again in excellent
agreement, and for our choice of $\gamma$, $T_{0}$ and $k_F$ (1.3,
$1.7\times10^4$ K, 6.5 Mpc$^{-1}$) the semi-analytical and
hydrodynamical simulations both match well with the observations of
\citet{Croft:2002p6751}.

We also compare the 1D flux PS computed from the lower resolution,
large volume simulation to the data. The dot-dashed curve in Figure
\ref{fig:FluxPS} displays the 1D PS computed from the unprocessed
spectra, which agrees well with the observational data and smaller box
simulations. Note, however, the PS in this case extends to much larger
spatial scales. The 1D flux PS for the \lya{} forest spectra that has
been processed to resemble low resolution data is shown by the dotted
curve in Figure \ref{fig:FluxPS}. On larger scales this matches the
power of the high resolution spectra from the 1 Gpc simulation well,
demonstrating that lower resolution leaves the large spatial scales
largely unaltered, as is required for measurements of the \bao{}
scale. Note, however, the low resolution PS exhibits less power at
smaller scales as expected, and also flattens out at $k>0.01$ s
km$^{-1}$ due to noise.

As in the case of the flux PDF, the flux PS is sensitive to our choice
of free parameters. In particular, the shape of the flux PDF is highly
sensitive to the smoothing scale $k_{F}$ (for $k>0.01$ s
km$^{-1}$). For a smaller $k_{F}$, the power on small scales decreases
as the underlying density field becomes smoother. The flux PS is also
sensitive to both the slope of the temperature-density relation
$\gamma$ and the temperature $T_{0}$, and the power decreases on small
scales for both an increasing $\gamma$ and temperature $T_{0}$ at
$z=3$ \citep[see also][]{Viel:2004p7045}. However, in this work our
main goal is to demonstrate that our simulations provide a model of
the \lya{} forest which is adequate for extracting the \bao{}
signature. The broad agreement with the observational data and
previous modelling suggests this is indeed the case.

\section[Extracting the BAO signal from the Ly-alpha forest]{Extracting the \bao{} signal from the $\bmath{\lya{}}$ forest} \label{Sec:Recovery}

The broad agreement between our semi-analytic simulations and the
data, as well as the excellent agreement of our models with our
reference hydrodynamical simulation, gives us confidence that our
semi-analytical simulations provide a good representation of the
\lya{} forest on large scales.  We therefore now proceed to extract
the \bao{} scale from our large scale simulations of the $z\sim3$
\lya{} forest.

\subsection[Mock data set]{Mock data set}

We generate mock data sets using spectra sub-samples selected at
random from a total sample of 100,000 lines of sight drawn parallel to
the box boundaries of our 1 Gpc simulation volume.  The total
simulation volume of 1 Gpc$^{3}$ corresponds to a survey area of $\sim
79\rm~deg^{2}$ at $z=3$.  We make the approximation that all
background sources are at the redshift of the simulation
(\textit{i.e.} $z=3$), the sight-lines are parallel, and that the spectra all have the same usable
pathlength (\textit{i.e.} the distance between the quasar rest frame
\lya{} and Ly$\beta$ transitions minus $3000$ km s$^{-1}$ to account
for the proximity effect). Note that because our method allows us to
generate many simulations using different realisations for the density
field at various redshifts quickly and efficiently, we may easily
extend the volume of a mock survey data set as required. Indeed much
larger survey volumes will be required in practice to extract the
\bao{} signature from observational data
(e.g. \citealt{McDonald:2007p6752}).

\subsection[Reconstructing the 3D PS from Ly-alpha spectra]{Reconstructing the 3D PS from $\bmath{\lya{}}$ spectra} \label{section:PSregen}

In practice, as there are only ever a limited number of skewers
(quasar sight-lines) drawn through a survey volume, we cannot directly
measure the full 3D \lya{} flux PS from the data. In order to estimate
the full 3D \lya{} flux PS from our simulations we must therefore
reconstruct the true 3D \lya{} flux PS from the 3D \lya{} PS computed
from the individual sight-lines, minus a weighted term which
introduces aliasing-like noise to the analysis. We adopt the approach
of \citet{McDonald:2007p6752} and use the following expression for the
reconstruction:

\begin{equation} \label{eq:reconstruction}
P_{F,\rm{true}}(\bmath{k}) = P_{F,\rm{box}}(\bmath{k}) - P_{F,\rm{1D}}(\bmath{k}_{\parallel})P_{2D,\rm{w}}(\bmath{k}_{\perp}) - P_{\rm{N}}.
\end{equation}

\noindent
Here we measure the full 3D PS, $P_{F,\rm{bo}x}(\bmath{k})$, of our
simulation volume using only the information given by the individual
lines-of-sight in the mock survey. We then subtract off a `3D' PS
generated by the multiplication of a 2D weighting PS,
$P_{2D,\rm{w}}(\bmath{k}_{\perp})$, indicating the positions of the
\lya{} spectra and the 1D \lya{} flux PS,
$P_{F,\rm{1D}}(\bmath{k}_{\parallel})$, measured from the synthetic
\lya{} spectra along the line-of-sight. Finally, we also subtract off
the noise PS, $P_{\rm{N}}$, which is generated from the noise along
the line-of-sight multiplied by the 2D weighting PS. After completion
of this reconstruction process we then spherically average the
resulting PS. We bin the spherically averaged 3D PS using concentric
spheres of radii equal to multiples of the Nyquist frequency. Since
the Fourier modes are discrete the position of each $k$-bin is
calculated by taking the average of all $k$ values which fall in each
concentric sphere. This binning strategy is to ensure we do not
introduce a shift in the \bao{} signal that can affect the recovered
value of the \bao{} scale.

We complete the reconstruction step defined by equation
(\ref{eq:reconstruction}) on our mock data sets from two simulations; one
using the matter PS including baryonic oscillations and another with
the smoothed reference PS. By taking the ratio of these two
reconstructed 3D \lya{} flux PS, we extract the resulting \bao{}
signature. Note that in this work we have simply given each individual
line-of-sight the same weighting in the reconstruction (\textit{i.e.}
one for each \lya{} spectrum and zero for no \lya{}
spectrum). Realistically, each weighting will vary according to the
quality of the \lya{} forest spectra
\citep{McDonald:2007p6752,McQuinn:2011p10709}.

\begin{figure*}
	\begin{center}
		\includegraphics[trim = 2cm 1cm 2cm 1cm, scale =
                  0.33]{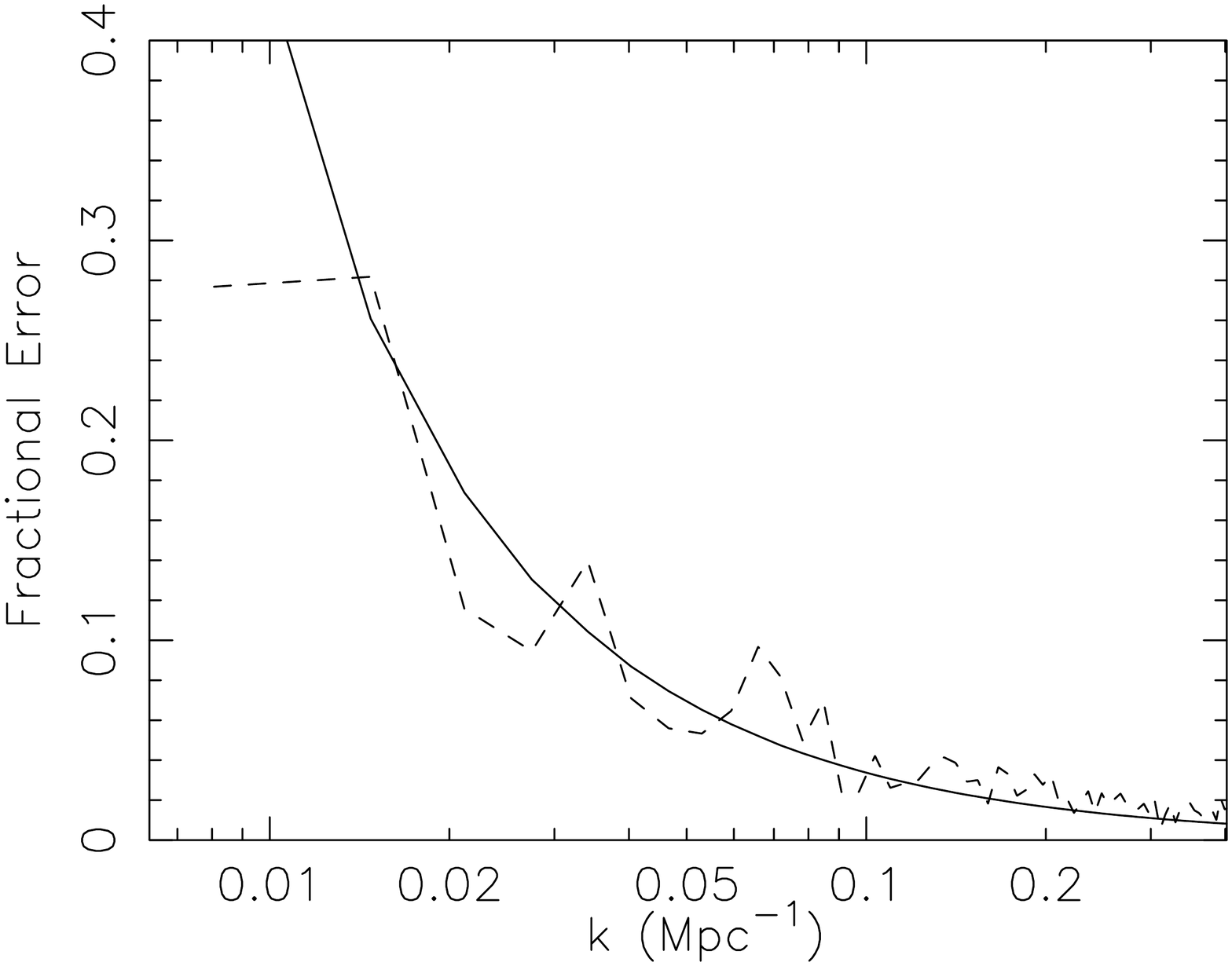}
                  \includegraphics[trim = 1cm 1cm 2cm 1cm, scale =
                  0.33]{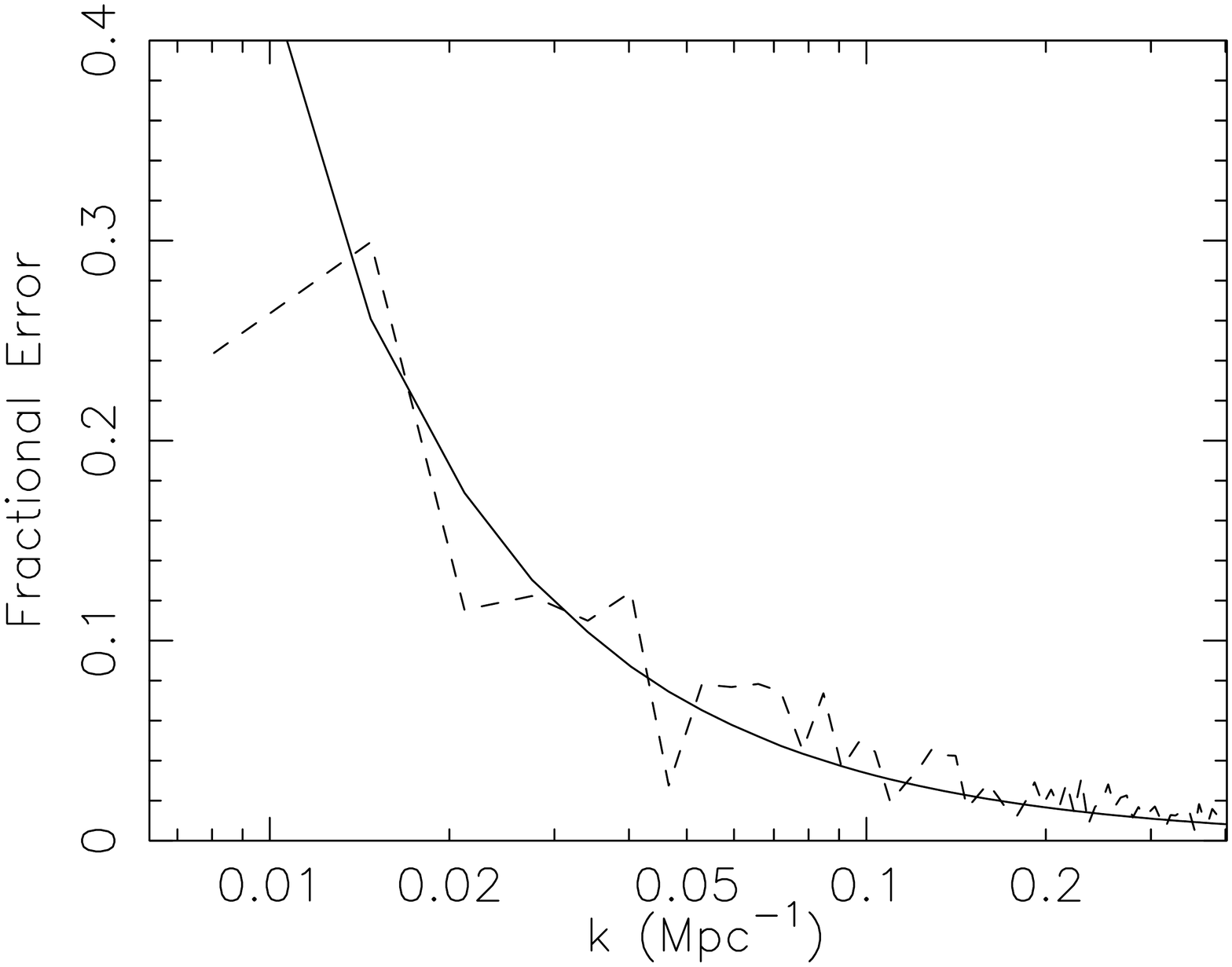}
                  \end{center}
                  \caption{Comparison of the measured cosmic variance
                    from nine different $1\rm\,Gpc^{3}$ volume
                    simulations (dashed curves) to the theoretical
                    expression given by equation (\ref{eq:CosVareq}) (solid
                    curves). \textit{Left panel:} The fractional error
                    on the measurement for the matter power
                    spectrum. \textit{Right panel:} The fractional
                    error on the \lya{} flux power spectrum.}
                  \label{fig:CosmicVariance}
                  \end{figure*}
                  
                  \begin{figure*}
	\begin{center}
		\includegraphics[trim = 2cm 1cm 2cm 1cm, scale =
                  0.65]{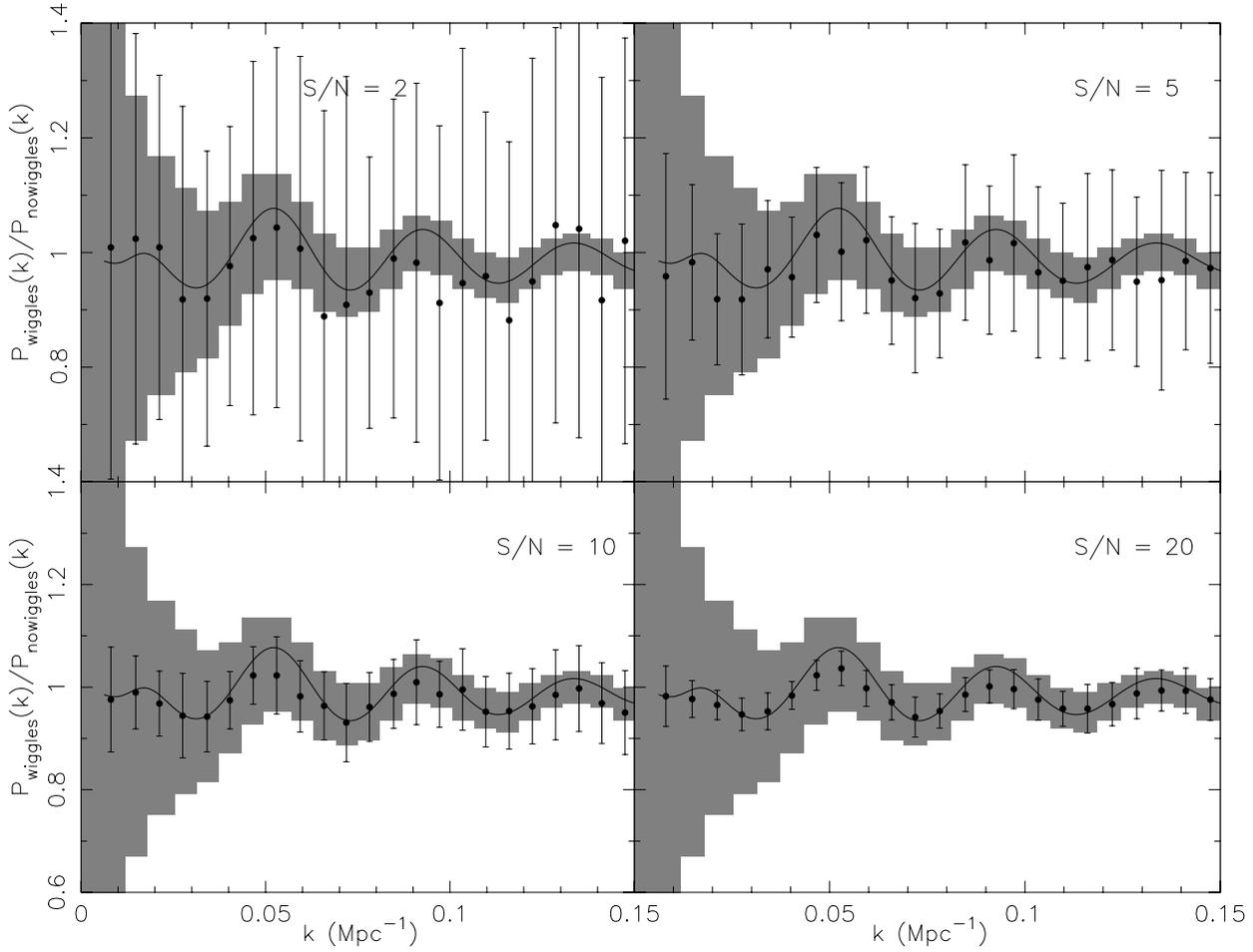}
	\end{center}
        \caption{The data points with 1-$\sigma$ error bars display
          the \bao{} signature recovered from the 3D \lya{} flux PS
          generated from a mock \lya{} dataset containing 1200 (at
          $\sim 15$ per square degree) lines of sight for a
          $79\rm~deg^{2}$ survey (see text for details). Clockwise
          from the top left, the recovered \bao{} signature is
          extracted from spectra constructed with S/N ratios of 2, 5,
          20 and 10. The error bars for varying S/N are generated
          using the Monte-Carlo approach described in Section
          \ref{section:ShotNoise}. For comparison, the solid curve in
          each panel is the expected \bao{} signature generated from
          the ratio of the input linear dark matter power spectra with
          and without the baryon oscillation features. The shaded
          region, which is the same in each panel, displays an
          estimate of the cosmic variance error using equation (\ref{eq:CosVareq}) for our mock survey volume
          of $1\rm\,Gpc^{3}$.}
\label{fig:Largesimwiggles}
\end{figure*}
   
\begin{figure*}
	\begin{center}
	\includegraphics[trim = 2cm 1cm 2cm 1cm, scale =
                  0.65]{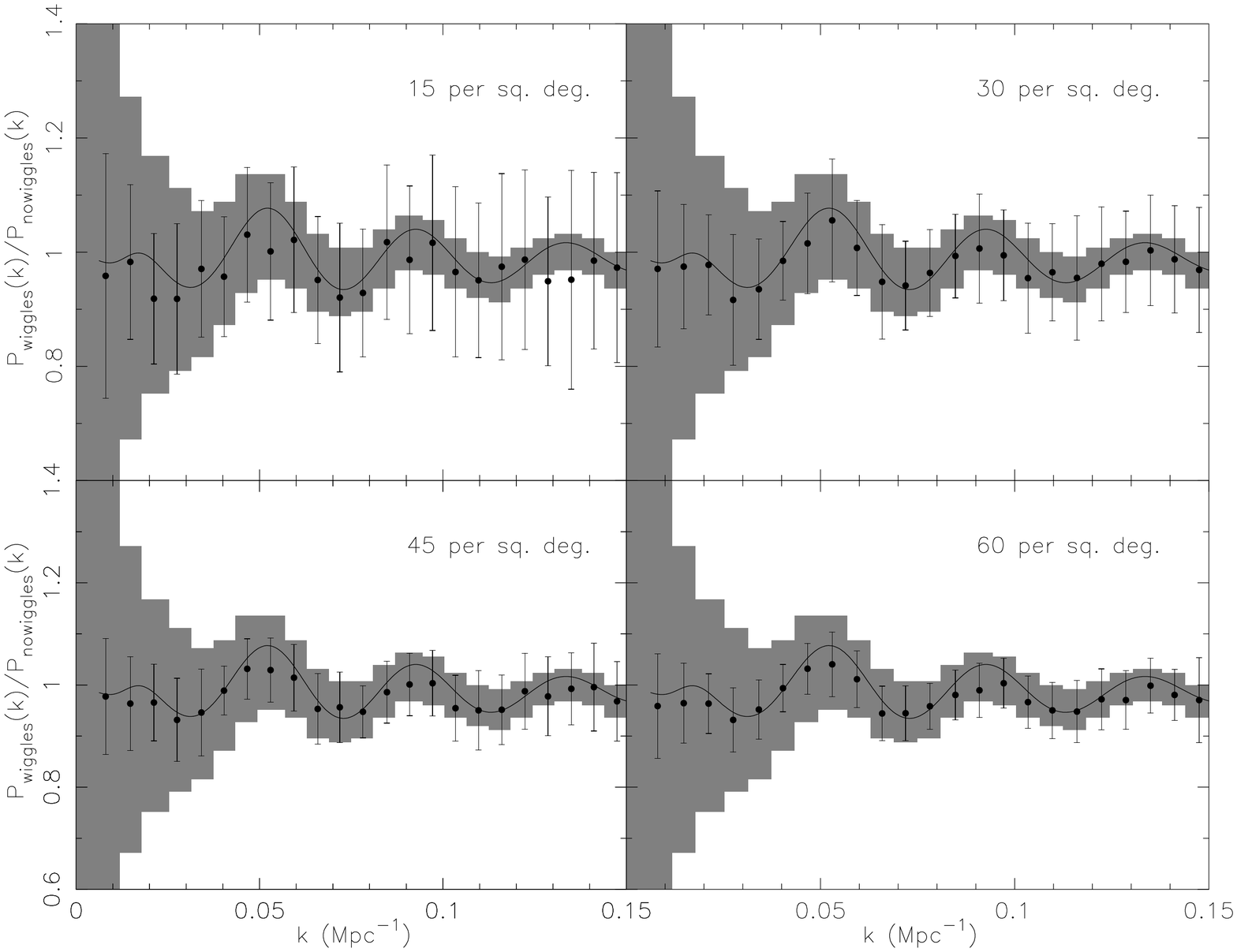}
	\end{center}
	\caption{Comparison of the Monte-Carlo generated shot noise
          errors (error bars) and the cosmic variance errors (shaded
          region) for varying background source density and a fixed
          S/N of $5$ per pixel. The Monte-Carlo shot noise errors are
          generated following the outline in Section
          \ref{section:ShotNoise}.  Clockwise from the top left, the
          recovered \bao{} signature is extracted from sight-line
          densities of $15$, $30$, $60$ and $45\rm ~deg^{-2}$.}
\label{fig:DiffSightlineDensity}
\end{figure*}

\subsection[Measurement uncertainties]{Measurement uncertainties}

There are two contributions to the uncertainty on the reconstructed
BAO signal we consider here; cosmic variance and shot noise. We shall
estimate the former by using eight further $1\rm\, Gpc^{3}$
simulations using different random seeds to generate the initial
conditions. For the latter we shall adopt a Monte-Carlo error bar
estimate.

\subsubsection[Cosmic variance]{Cosmic variance}
 
In Figure \ref{fig:CosmicVariance} we estimate the fractional error on
the power spectrum measurement due to cosmic variance. We use nine
different $1\rm\, Gpc^{3}$ simulations to generate the 3D matter (left
panel) and \lya{} forest power spectra (right panel) for each
realisation. The \lya{} power spectra from each simulation box is
reconstructed using $20,000$ noiseless spectra at a density of
$\sim250\rm \,deg^{-2}$. The shot noise error due to under-sampling
becomes negligible for this artificially high background source density. Using
these simulations we then estimate the cosmic variance error by
measuring the 1-$\sigma$ variations in the power spectra across the 9
different boxes, shown as the dashed curves in Figure
\ref{fig:CosmicVariance}. These are compared to the theoretical
expression, displayed as the solid curves in each panel, given by
equation (2) in \citet{Blake:2003p6749},
\begin{equation} \label{eq:CosVareq}
\left(\frac{\sigma_{P}}{P}\right)^{2} = 2 \frac{(2\pi)^{3}}{V}\frac{1}{4\pi k^{2} \Delta k},
\end{equation}
which gives the error on a power spectrum measurement averaged over
spherical $k$-bins of width $\Delta k$. For both the matter and \lya{}
forest power spectra the theoretical expression is in good agreement
with our estimated cosmic variance errors except at the largest
scales, $k\la 0.015$ Mpc$^{-1}$. Hence we choose to use equation (\ref{eq:CosVareq}) for cosmic variance in the
remainder of this work.
                
\subsubsection[Shot noise]{Shot noise} \label{section:ShotNoise}

We adopt a Monte-Carlo approach for estimating the shot noise on our
extracted power spectrum.  We randomly sample our 1 Gpc$^{3}$ \lya{}
simulation (containing 100,000 lines-of-sight over $\sim
79\rm~deg^{2}$) and generate subsamples of 1200 spectra, yielding
$\sim 15$ spectra per square degree. For each randomly generated
subsample, we complete the PS reconstruction process outlined in
Section \ref{section:PSregen}. We perform this procedure 100 times
using spectra which have four different signal-to-noise ratios
(S/N=$2,\,5,\,10$ and $20$), and estimate the full covariance matrix
and 1-$\sigma$ error for each $k$-bin.

\subsubsection[Error bar estimates]{Error bar estimates}

In Figure \ref{fig:Largesimwiggles} we compare the reconstructed
\bao{} signature generated from our mock \lya{} forest data to the
input \bao{} signature generated from the matter PS for different
assumptions regarding the S/N ratio of the data.  The spectra have
been sampled at a density of $15$ quasars per square degree
(corresponding to the planned density of BOSS targets).  We also
compare our Monte-Carlo error bars to the expected cosmic variance
error due to survey volume (shaded region) computed using equation
(\ref{eq:CosVareq}).

For S/N ratios of 2 and 5 per pixel the Monte-Carlo estimates of the
errors are larger than the expected cosmic variance error for a $\sim
79 \rm~deg^{2}$ survey area. As the S/N is increased to $> 10$ the
recovery of the \bao{} signature becomes limited by cosmic
variance. We reiterate that the simulations used a volume of 1
Gpc$^{3}$ and that the error bars are proportional to the inverse
square root of the survey area. We use a volume much smaller than a
survey would require in order to illustrate the size of the error
bars.

In Figure \ref{fig:DiffSightlineDensity}, we instead compare the
Monte-Carlo shot noise errors to the cosmic variance error estimates
while varying the background source density and assuming a fixed S/N
ratio of $5$ per pixel.  The line-of-sight densities are 15, 30, 45 and
finally 60 quasars per square degree; background source densities
larger than 45 per square degree are cosmic variance limited for our
mock survey area of $79\rm~deg^{2}$.

\subsection[Recovery of the BAO scale]{Recovery of the \bao{} scale}

We now quantify the recovery of the \bao{} signature from our
simulations by fitting the recovered 3D \lya{} flux PS with a function
dependent on the characteristic \bao{} scale length. Following the
approach of \citet{Blake:2003p6749}, we assume a simple two parameter
decaying sinusoidal model for the ratio of the 3D PS with baryonic
oscillations and the smooth reference PS. The functional form of the
assumed fitting function is:

\begin{eqnarray} \label{eq:twoparameter}
\frac{P_{F}(k)}{P_{F,\rm{r}}(k)} &=& 1.0 + Ak\rm{exp}\left[-\left(\frac{\it{k}}{0.1\:\it{h}\:\rm{Mpc}^{-1}}\right)^{1.4}\right]
\nonumber \\
& & \times \: \rm{sin}\left(\frac{2\pi \it{k}}{\it{k_A}}\right),
\end{eqnarray}

\noindent
where $P_{F,\rm{r}}$ is the smoothed reference PS with no baryon
oscillations, $A$ is an arbitrary normalisation constant and $k_{A}$
is the characteristic \bao{} scale in Fourier space (where the
characteristic \bao{} scale $s_{A} = 2\pi/k_{A}$). To determine the
two unknown parameters we perform a $\chi^{2}$ minimisation for the
ratio of the PS obtained from our simulation to the function in
equation (\ref{eq:twoparameter}), such that:

\begin{eqnarray} \label{eq:chi2}
\chi^{2}(\bmath{p}) &=& \sum_{i=1}^{n_k}\sum_{j=1}^{n_k}{\textbfss{C}_{ij}}^{-1}[P_{\rm{ratio}}(k_{i}) - P_{\rm{ratio,fit}}(\bmath{p},k_{i})] \nonumber \\
& & \times[P_{\rm{ratio}}(k_{j}) - P_{\rm{ratio,fit}}(\bmath{p},k_{j})],
\end{eqnarray} 
\noindent
where $\bmath{p}$ contains the parameters for the fitting formula ($A$
and $k_{A}$), and ${\textbfss{C}_{ij}}^{-1}$ is the inverse of the
covariance matrix generated from our Monte-Carlo procedure. Here
$P_{\rm{ratio}}(k)$ is the ratio of the two simulated PS (with and
without the baryon oscillations) and $P_{\rm{ratio,fit}}(\bmath{p},k)$
is given by equation (\ref{eq:twoparameter}). We restrict the range of
our $\chi^{2}$ minimisation to wavenumbers below $k = 0.25$
Mpc$^{-1}$, where the summation index denotes the summation over $k$
space bins in our simulated PS.

\begin{table*}
\begin{tabular}{@{}lcccc}
\hline

Number Density & S/N & $s_A$ ($2\pi/k_{A}$) & $\Delta s_A$ & $\Delta s_A$ (2000 deg$^{2}$)\\
(deg$^{-2}$) &  & (comoving Mpc) & (per cent)  & (per cent)\\
\hline

15 & 5 & 150.7 & 12.70 & 2.53\\
 & 10 & 157.4 & 8.51 & 1.70\\
 & 20 & 154.2 & 2.64 & 0.52\\
 & Noiseless & 153.1 & 0.63 & 0.13\\
\hline

45 & 5 & 155.7 & 5.73 & 1.14\\
 & 10 & 152.0 & 2.41 & 0.48\\
 & 20 & 151.9 & 1.23 & 0.24\\
 & Noiseless & 152.4 & 0.30 & 0.06\\
\hline

\end{tabular}
\caption{The recovered best fitting values for the \bao{} scale,
  $s_{A}$, along with the fractional 1-$\sigma$ errors obtained by
  applying the $\chi^{2}$ minimisation described by equation
  (\ref{eq:chi2}) to the simulated data displayed in Figure
  \ref{fig:Largesimwiggles}.  Results are shown for a background
  source density of $15$ and $45\rm~deg^{-2}$ and for S/N ratios of
  $5$, $10$ and $20$.  The $\chi^{2}$ minimisation is performed using
  only the Monte-Carlo shot noise error estimates. The final row also
  provides the parameters recovered from noiseless data. For
  comparison, the input value is $s_{A}=152.5$ comoving Mpc. The final column
  contains the fractional error on the \bao{} scale after scaling our
  simulation result to a 2000 square degree survey area
  (\citealt{McDonald:2007p6752}).}
 \label{BAOtable}
\end{table*}

\begin{table*}

\begin{tabular}{@{}lcccc}
\hline

Number Density & S/N & $s_A$ ($2\pi/k_{A}$ Mpc) & $\Delta s_A$ & $\Delta s_A$ ($10^{4}$ deg$^{2}$)\\
(deg$^{-2}$) &  & (comoving Mpc) & (per cent) & (per cent)\\
\hline

15 & 5 & 150.7 & 15.51 & 1.38\\
 & 10 & 152.3 & 14.59 & 1.30\\
 & 20 & 152.0 & 6.28 & 0.56\\
 & Noiseless & 153.5 & 3.75 & 0.33\\
\hline

45 & 5 & 155.6 & 9.12 & 0.81\\
 & 10 & 152.6 & 5.66 & 0.50\\
 & 20 & 151.3 & 4.46 & 0.40\\
 & Noiseless & 151.1 & 4.24 & 0.38\\
\hline

\end{tabular}
\caption{As for as Table 1, except now showing the recovered best
  fitting parameters for the \bao{} scale, $s_A$, using both the
  Monte-Carlo shot noise errors and the cosmic variance summed in
  quadrature.  The final column contains the fractional error on the
  \bao{} scale after scaling our simulation result to a $10^{4}$ square
  degree survey area.}
 \label{BAOtable2}
\end{table*}

The best fit values for $s_{A} = 2\pi/k_{A}$ are summarised in
Table~\ref{BAOtable} for various S/N ratios (including the idealised
case of noiseless data) assuming a background source density of either
$15$ or $45\rm~deg^{-2}$. The 1-$\sigma$ relative errors on $s_{A}$
are again estimated by using 100 Monte-Carlo subsamples of mock
$\lya{}$ data. For each subsample we performed the $\chi^2$
minimisation and used the distribution of recovered values to estimate
the 1-$\sigma$ relative error, $\Delta s_{A}$.  The results are shown
for a 1 Gpc$^{3}$ volume with an area of $79\rm~deg^{2}$, and are
also shown after scaling to $2000\rm~deg^{2}$ (assuming $\Delta
s_{\rm A}$ is proportional to the inverse square root of the survey
area). We do not quote the best fit parameters and their associated
errors for S/N$=2$, as we could not recover the characteristic \bao{}
scale at all for our small simulated volume. Even for S/N$=5$, the
recovery of the \bao{} signal is fairly poor.  However, this is to be
expected for the small survey area used here, and an increase in the
survey area or S/N would improve the fractional error on the recovered
\bao{} scale considerably (\citealt{McDonald:2007p6752}). We
nevertheless find that we are able to recover the \bao{} scale from
our 1 Gpc$^{3}$ simulations to within a few percent for S/N$>5$.

In Table~\ref{BAOtable2} we again provide the best fit values for
$s_{\rm A}$, but we instead perform the $\chi^2$ minimisation using
the Monte-Carlo shot noise errors added in quadrature with the cosmic
variance errors from equation (\ref{eq:CosVareq}). We find that for 15
quasars per square degree, increasing the signal to noise continues to
reduce the fractional error on the \bao{} scale.  However for 45
quasars per square degree the fractional error saturates for S/N $>$
10, indicating the mock survey is cosmic variance limited for these
parameters.  

As a consistency check, we also obtain the recovered \bao{} signal
from noiseless spectra. We performed the same 100 subsample estimation
of the error bars both ignoring the estimated sample variance (Table
\ref{BAOtable}) and including the sample variance (Table
\ref{BAOtable2}). We recover a \bao{} scale of 152.4 comoving Mpc (shot noise error only, 45 \rm$\deg^{-2}$), compared
with the \bao{} scale of the input PS which was 152.5 comoving Mpc. This
confirms that our semi-analytical simulations provide a reasonable
description of correlations in the density field on large scales. 

We note that the two parameter fitting formula used in equation
(\ref{eq:twoparameter}) does not provide a perfect description of our
simulated data. There is a small reduction in the amplitude of the
power spectrum toward small scales which arises following the mapping
and smoothing of the initial linear density field described in Section
2. This means the oscillations do not occur exactly around one mean
value as expected in the fitting formula, and this affects the
recovery of the arbitrary constant, $A$, in equation
(\ref{eq:twoparameter}). Importantly, however, this does not affect the
recovery of the \bao{} scale from our simulations.

\subsection[Comparison to other work]{Comparison to other work}

Several other authors have recently described simulations designed to
recover the \bao{} signature from the \lya{} forest. It is therefore constructive to compare the results presented in this work to
the approaches taken by other studies using our results from Table
\ref{BAOtable}. Firstly, \citet{McDonald:2007p6752} used analytical
arguments to predict that one should be able measure the radial and
transverse distance scales used for \bao{} measurements to within a
fractional error of $\sim$1.4 per cent for ${\rm S/N}=1.8$ per pixel,
$\sim$40 quasars per square degree and a survey area of 2000 square
degrees.  For a S/N of 5 per pixel over $\sim$79 square degrees for
$\sim$45 quasars per square degree, we obtain a fractional error on
the \bao{} scale of $\sim$5.7 per cent. Scaling our survey area to
2000 square degrees (assuming the error scales as the inverse square
root of the survey area) we would expect a fractional error of
$\sim$1.1 per cent (final column Table \ref{BAOtable}), consistent
with the results of \citet{McDonald:2007p6752}.

Large N-body simulations were used by \citet{Slosar:2009p5710} to
measure the correlation function and also the cross-PS, rather than
the 3D PS. Each dark matter N-body simulation used by
\citet{Slosar:2009p5710} contained $3000^{3}$ particles in a box of
size 1500 $h^{-1}$Mpc, which does not fully resolve the Jeans scale.
However, the authors argue this should not affect the power on the
acoustic scale. They conclude that the correlation function provides a
better method for recovery of the \bao{} signal compared to both the
3D flux PS and the cross-PS, which is consistent with the approach
taken by \citet{White:2010p5647}. \citet{Slosar:2009p5710} fit for the
\bao{} peak at $z=2.5$ using $\sim$56 quasars per square degree over a
total area of $\sim400$ square degrees. For noiseless spectra we find
a fractional error on the \bao{} scale of 0.30 per cent compared to
0.53 per cent from \citet{Slosar:2009p5710}. The survey area of
\citet{Slosar:2009p5710} is $\sim$5 times larger than ours, implying
an error that should be $\sim$2.2 times smaller than our error. The
origin of this difference is unclear, but may be in part due to the
use of the diagonal covariance by \citet{Slosar:2009p5710}, as opposed
to the full covariance matrix we use in our best-fit parameter
estimation.

Finally, we compare our results to those of \citet{McQuinn:2011p10709}
who provide sensitivity estimates for large \lya{} forest surveys.
Although a direct comparison here is less straightforward, using their
table 4, a background source density of $\sim 15\rm~deg^{-2}$
corresponds to an $\bar{n}_{eff}= 1.4\times 10^{-3}$ Mpc$^{-2}$ at
$z=2.5$. However, at $z=2.5$ for S/N = 2, the effective number density varies from the true number density by roughly 0.59 (their table 3) and so $\bar{n}_{eff}\sim 0.8\times 10^{-3}$ Mpc$^{-2}$. From figure 8 in \citet{McQuinn:2011p10709}, this gives the
fractional precision in the angular diameter distance and Hubble
expansion of $\sim 2 - 3$ per cent for a $\sim10^4$ square degree
survey volume at $z=2.5$. Scaling our $z=3$ results
for the fractional precision of the \bao{} scale assuming S/N of 2
(Section \ref{Sec:Stats}), a survey area of $10^{4}\rm\,deg^{2}$ and
15 quasars per square degree, we find a fractional precision of
$\sim2.5$ per cent on the \bao{} scale. \citet{McQuinn:2011p10709} also find that the amount of information obtained from a quasar is maximised for S/N $\sim$5-10, consistent with our findings in Figure \ref{fig:Largesimwiggles}.
\section{Approximate scaling relations for the fractional error} \label{Sec:Stats}

To summarise our results we provide approximate scaling relations for the
expected recovered fractional error on the \bao{} scale as a function
of both survey S/N and quasar number density.  We approximate this
using a simple power law expression,

\begin{equation}
\rm{fractional~error~(per~cent)} = ax^b\left(\frac{\textrm{Survey area}}{79~\textrm{deg$^{2}$}}\right)^{-1/2},
\label{bestfit}
\end{equation}
where $\textrm{x}$ is either S/N (Table \ref{SNfit}) or quasar number density
(Table \ref{densfit}) and perform a least squares fit on our
simulation results. We provide the best fit parameters $a$ and $b$ for
the fractional error fit for the shot noise errors alone, as well as
the shot noise added in quadrature with the cosmic variance.  The
fractional error also scales as the inverse square root of the survey
area.

\subsection[Scaling constraints to a BOSS-like survey]{Scaling constraints to a BOSS-like survey}

Using the best fit parameters from our model for the \bao{}
scale from Table \ref{BAOtable2}, we can estimate the accuracy of
forthcoming large volume \bao{} \lya{} forest surveys. BOSS
anticipates the detection of 150,000 quasars in a total survey area of
$10^{4}\rm~deg^{2}$, with S/N = 5 and a source number density of
$15\rm~deg^{-2}$.  Scaling our simulation results from Table
\ref{BAOtable2}, we anticipate a detection of the \bao{} scale to
within $\sim$1.4 per cent including cosmic variance.

\begin{table}
\begin{center}
\begin{tabular}{@{}lccc}
\hline
 Error & Number Density & $a $& $b$\\
 & (sq. deg.) &  & \\
\hline
 Shot noise & 15 & 52.85 & -0.91 \\
 & 45 & 26.50 & -1.04 \\
\hline
Shot noise & 15 & 41.34 & -0.54\\
$+$cosmic variance & 45 & 27.55 & -0.66 \\
\hline

\end{tabular}
\caption{Best fit parameters for $a$ and $b$ in equation
  (\ref{bestfit}) for the fractional error on $s_{A}$ for varying
  S/N (the models used to obtain the scaling fit have S/N=2, 3.5, 5,
  7.5, 10, 15 and 20). Scalings are given for background source
  densities of both 15 and 45 quasars per square degree, and a total
  survey area of $79\rm~deg^{2}$. \textit{Top row}: Fit parameters
  for Monte-Carlo shot noise error bars only. \textit{Bottom row}: Fit
  parameters for shot noise and cosmic variance errors added in
  quadrature.}
 \label{SNfit}
 \end{center}
\end{table}

\begin{table}
\begin{center}
\begin{tabular}{@{}lccc}
\hline
 Error & S/N & $a$ & $b$\\
\hline
 Shot noise & 5 & 99.12 & -0.78 \\
 & 10 & 140.39 & -1.14\\
\hline

Shot noise & 5 & 76.16 & -0.61\\
$+$cosmic variance & 10 & 53.35 & -0.55 \\
\hline

\end{tabular}
\caption{Best fit parameters for $a$ and $b$ in equation
  (\ref{bestfit}) for the fractional error on $s_{A}$ for varying
  quasar sightline density (the models used to obtain the scaling fit have
  5, 10, 15, 22.5, 30, 45 and 60 $\rm deg^{-2}$).  Scalings are given
  for both S/N = 5 and 10, and a total survey area of
  $79\rm~deg^{2}$. \textit{Top row}: Fit parameters for Monte-Carlo
  shot noise error bars only.  \textit{Bottom row}: Fit parameters for
  shot noise and cosmic variance errors added in quadrature.}
 \label{densfit}
 \end{center}
\end{table}

\section[Conclusion]{Conclusion} \label{Sec:Conclusion}

A series of recent studies have used large volume, high resolution
N-body and hydrodynamical simulations to study the detection of \bao{}
in forthcoming large volume \lya{} forest surveys (such as BOSS). One
limitation of these simulations is the large computational cost
required for a simulation of sufficient volume. On the other hand, in order
to study the systematics involved in the detection of the \bao{}
signal, it will be critical to run many simulations to fully probe
parameter space. In this work we have demonstrated that a
semi-analytical model utilising a density field calibrated against a
hydrodynamical simulation can be used to produce very large volume,
high resolution simulations at a fraction of time and computational
cost, and at a reasonable level of accuracy. We find good quantitative
agreement between the semi-analytical model and the hydrodynamical
simulation for a range of observables. In particular, we are able to
reproduce one and two-point statistics (the flux PDF and PS), which are
in reasonable agreement with observational data. We stress that in this work we assume a constant redshift, $z=3$, whereas one expects that the accuracy of the recovered \bao{} scale with be redshift dependant, and hence our results will differ slightly for different redshifts.

We used our model to generate mock \lya{} forest data sets drawn from
a 4096$^3$ 1 Gpc$^3$ simulation volume. We demonstrated we are able to recover
the \bao{} signal through reconstruction of the 3D \lya{} PS
\citep{McDonald:2007p6752}. We also recover the characteristic
\bao{} scale length by applying a $\chi^{2}$ minimisation with a
simple two parameter fitting function \citep{Blake:2003p6749}. We
used Monte-Carlo realisations of the \lya{} forest data to estimate
the $1-\sigma$ uncertainties on the PS due to shot noise for varying
S/N and background source densities, and include an estimate of cosmic variance error from our nine 1 Gpc$^3$ simulations. Our mock surveys of
$\sim$ 15 quasars per square degree over $\sim$ 79 square degrees with
S/N = (5, 10, 20, $\infty$) yield relative errors of (12.7, 8.51,
2.64, 0.63) per cent and for $\sim$ 45 quasars per square degree
(5.73, 2.41, 1.23, 0.30) per cent on the recovered \bao{} scale. The
accuracy to which we can recover the \bao{} scale with square root of
the inverse volume, and are consistent with the predictions presented by
\citet{McDonald:2007p6752}, \citet{Slosar:2009p5710} and
\citet{McQuinn:2011p10709}.

Using the results of our mock \lya{} analysis, we anticipate that for
a S/N = 5 with 15 quasars per square degree for a BOSS-like survey of
$10^{4}\rm~deg^{2}$, one should expect a fractional error of $\sim$
1.4 per cent on the \bao{} scale. We also provide simple
scaling relations for estimating the expected fractional error on the
\bao{} scale given the number density of quasars and the signal to
noise.

The method presented here enables generation of large scale
realisations of the IGM density field and \lya{} forest quickly and
efficiently. It is therefore ideal for investigating the key
systematics which will impact on \bao{} detection, such as errors in
the continuum shape and the effect of non-gravitational fluctuations
on the \lya{} forest, including large-scale temperature and ionisation
variations in the IGM at $z=3$.

\section*{Acknowledgments}

BG acknowledges the support of the Australian Postgraduate Award.  The
Centre for All-sky Astrophysics is an Australian Research Council
Centre of Excellence, funded by grant CE11E0090. We also thank Vincent
Desjacques for providing his low resolution \lya{} PDF data and Paul Geil
for providing the initial conditions generation code. We would also like to thank Chris Blake and Matt McQuinn
for comments on the draft manuscript.

\appendix
\section[GPGPU]{GPGPU}

In this appendix we describe the use of GPGPU (General Purpose
computing on Graphics Processing Units) programming in our
simulations. Our code has been implemented in three formats, single
core, parallelised multicore and a mix of parallel multicore and GPU
programming. The larger simulations used in this paper are performed
using both a parallel multicore and a single GPU. For all simulations
we use an Intel Xeon 2.00Ghz quad core CPU and a nVidia FX580 CUDA
enabled graphics card.

Over the last few years major steps have been taken in the
implementation of GPU programming into many astrophysical
applications. Implementations of current astrophysical simulations
with GPUs can report upward of a 10-100 factor speed up in
computational time \citep[see][and references
therein]{Fluke:2010p6029}.  One of the major problems for GPU programmers is how to
take full advantage of a GPU for solving computational problems. One
must maximise usage of on-chip resources, while allowing as much of
the calculation to be run without intervention from the host CPU. One
must also avoid data dependency, where a result at one point in the
data can impact on the outcome of a separate piece of data (i.e. data
must be as independent as possible). The transfer of data from the CPU
and GPU can also limit the effectiveness of the GPU programming
application, and is dependent on the details of the individual's
computer.

The goal of our simulations is to provide a model that can be used to
investigate the $\lya$ forest.  To accomplish this we only generate a
limited number of sightlines per simulation rather than the entire
density field (i.e. we generate our density field in Fourier space,
and only Fourier transform the number of sightlines required).  Our
semi-analytical model is perfectly suited for implementation on a GPU,
allowing us to quickly run mock survey simulations in less than a day.

\begin{figure}
	\begin{center}
		\includegraphics[trim = 0.5cm 2cm 0cm 2cm, scale = 0.3]{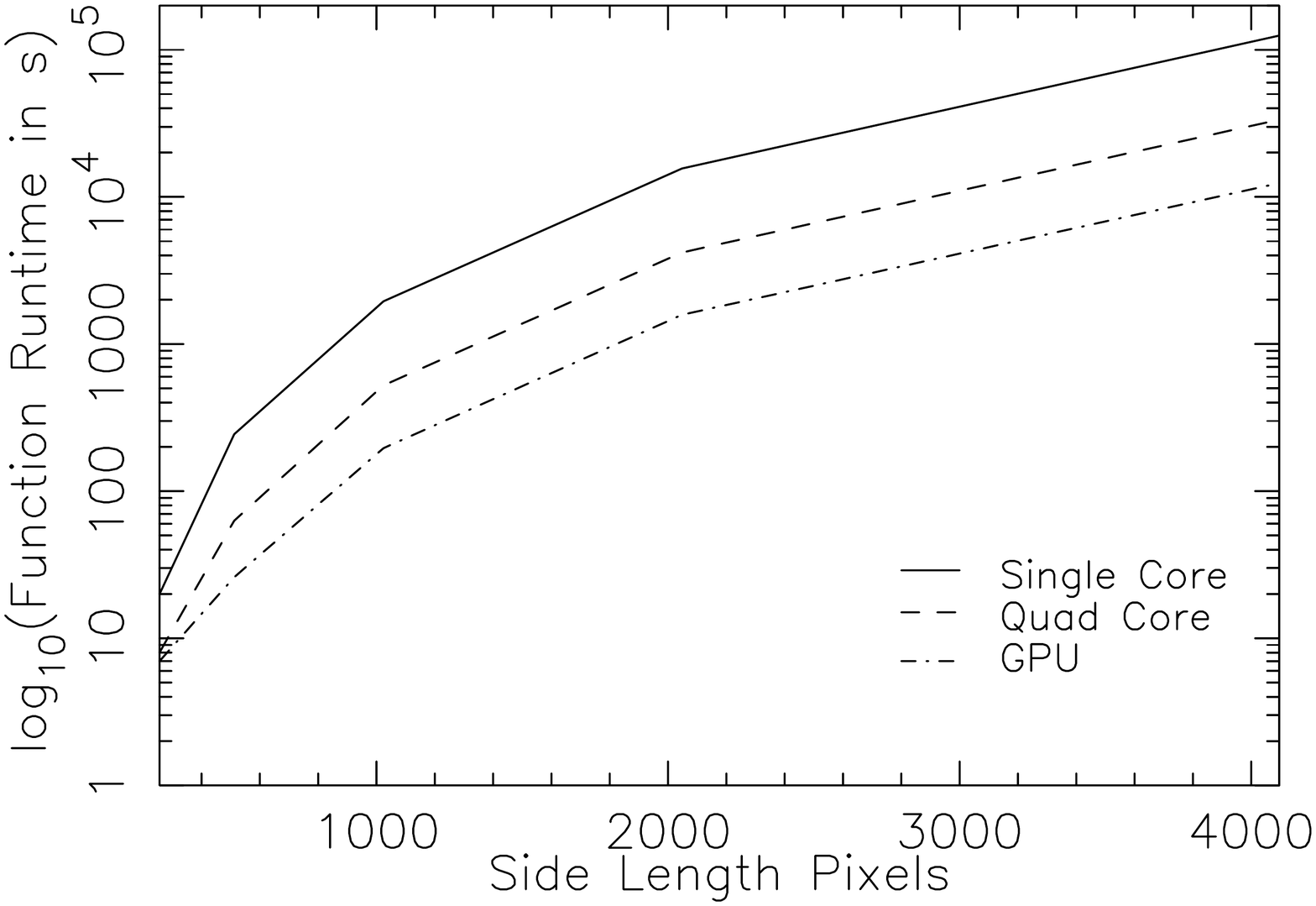}
	\end{center}
        \caption{Computation time for the generation of the density
          field in Fourier space as a function of the pixel number per
          side of a simulation cube. The solid curve shows the
          timings for a single core calculation, the dashed for quad
          core, and the dot-dashed curve is obtained using the
          GPU. The timing takes into account all required overhead
          such as data transfer between devices and memory
          allocation.}
\label{fig:denstiming}
\end{figure}

\begin{figure}
	\begin{center}
		\includegraphics[trim = 0.5cm 2cm 0cm 2cm, scale = 0.3]{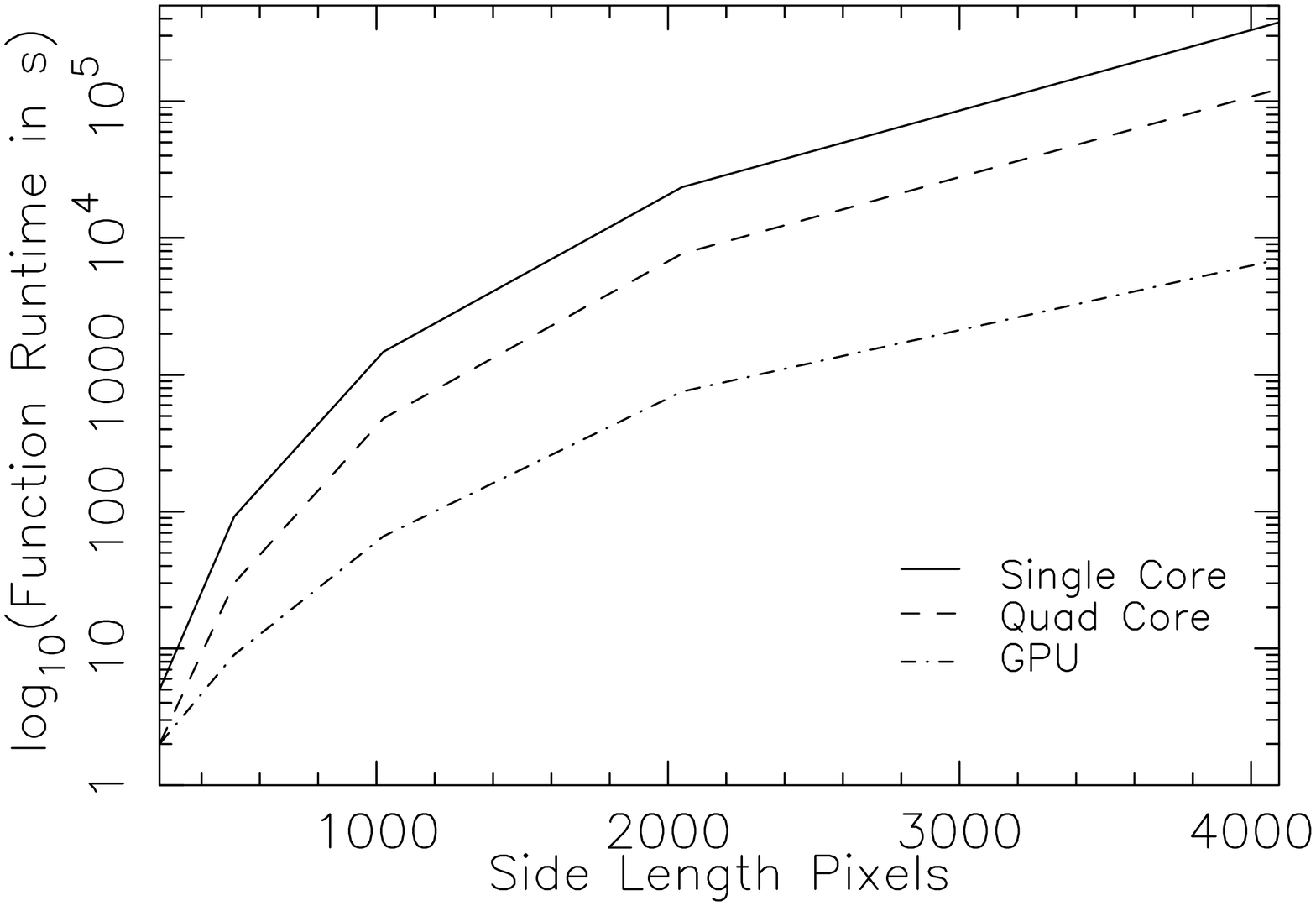}
	\end{center}
        \caption{Computation time for the generation of the 3D PS as a
          function of the pixel number per side of a simulation
          cube. The curves are as described for
          Figure~\ref{fig:denstiming}.}
\label{fig:PStiming}
\end{figure}

In Figures~\ref{fig:denstiming} and~\ref{fig:PStiming}, we show the
increase in performance gained by using a GPU for certain functions in
our simulation, relative to single and parallelised quad core
implementations. Figures \ref{fig:denstiming} and \ref{fig:PStiming}
show the total runtime of the individual section of the code used to
generate the Fourier space density field and for calculating the
spherically averaged 3D PS. We scale up the number of pixels along the
length of the simulation cube in powers of two, from 256 to 4096. We
include in the timing all required overhead, such as data transfer
from CPU to GPU and memory allocation.

For calculation of the Fourier space density field we observe a
reduction in runtime by roughly a factor of 4 on moving from the
single to quad core implementation, with a further factor of 4 from
quad core to GPU. For the 3D PS, we observe a factor 3 decrease in
runtime from single to quad core, and a factor of 8 decrease from quad
to GPU. Although our the decrease in runtime is only a factor of
$\sim$4-8 (relative to the parallelised CPU) for our two chosen
processes, this is mainly due to the specific GPU used. Our GPU
contains only 1.12Ghz clock speed, with a maximum of 32 cores, whereas
top of the line GPGPUs allow up to as many as 440 cores with clock
speeds of 1.3Ghz. Implementation of our GPU enabled code onto one of
the newest devices would facilitate the 10-100 factor increase in computational speed.

A considerable amount of computational time is taken up by Fourier
transformation of our simulated data. Although not implemented in the
current version of our code, initial testing of the inbuilt FFT
libraries provided show an expected additional factor of $\sim$10
reduction in computation time, which would further reduce our total
computation time.

\section[Memory limitations when generating large simulation boxes]{Memory limitations when generating large simulation boxes}

In this work we have used our code to generate 4096$^3$ simulation
boxes on a desktop PC. However on a desktop PC we are memory limited
and cannot store the entire box in memory at once. We circumvent this
problem by using the natural symmetry of the density field in Fourier
space to break up our large simulation volume into 8 smaller
simulation volumes. However, even these 8 smaller simulation volumes
cannot be fully read into memory at one time, and hence we only read
into memory a small section of the volume at any one time. The
advantage of such an approach is that our code can be performed easily
on any desktop computer.

Instead of computing the 3D FFTs (Fast Fourier Transform) which would
require the full simulation to be read into memory, we again use the
symmetry of the density field and Fourier transform in 2D slices
across our large simulation volume (reading in the necessary 4 smaller
simulation volumes for each 2D slice). We then only Fourier transform
out the final 1D lines of sight that we have randomly generated
throughout the full simulation volume. This final step of only Fourier
transforming the final 1D line of sight is well suited for generating
mock \lya{} forest surveys.

\bibliography{PapersDraft1}

\begin{thebibliography}{60}
\expandafter\ifx\csname natexlab\endcsname\relax\def\natexlab#1{#1}\fi

\bibitem[{Becker {et~al.}(2011)Becker, Bolton, Haehnelt, \&
  Sargent}]{Becker:2011p7578}
Becker G.~D., Bolton J.~S., Haehnelt M.~G., Sargent W. L.~W., 2011, Monthly
  Notices of the Royal Astronomical Society, 410, 1096

\bibitem[{Beutler {et~al.}(2011)}]{Beutler:2011p10735}
Beutler F., {et~al.}, 2011, eprint arXiv, 1106, 3366, 18 pages, 17 figures, 3
  tables

\bibitem[{Bi(1993)}]{Bi:1993p69}
Bi H., 1993, Astrophysical Journal, 405, 479

\bibitem[{Bi \& Davidsen(1997)}]{Bi:1997p3607}
Bi H., Davidsen A.~F., 1997, Astrophysical Journal v.479, 479, 523

\bibitem[{Blake {et~al.}(2007)Blake, Collister, Bridle, \&
  Lahav}]{Blake:2007p6108}
Blake C., Collister A., Bridle S., Lahav O., 2007, Monthly Notices of the Royal
  Astronomical Society, 374, 1527

\bibitem[{Blake \& Glazebrook(2003)}]{Blake:2003p6749}
Blake C., Glazebrook K., 2003, The Astrophysical Journal, 594, 665

\bibitem[{Blake {et~al.}(2011)}]{Blake:2011p10710}
Blake C., {et~al.}, 2011, Monthly Notices of the Royal Astronomical Society,
  951

\bibitem[{Bolton \& Becker(2009)}]{Bolton:2009p4777}
Bolton J.~S., Becker G.~D., 2009, Monthly Notices of the Royal Astronomical
  Society: Letters, 398, L26

\bibitem[{Bolton {et~al.}(2010)Bolton, Becker, Wyithe, Haehnelt, \&
  Sargent}]{Bolton:2010p6750}
Bolton J.~S., Becker G.~D., Wyithe J. S.~B., Haehnelt M.~G., Sargent W. L.~W.,
  2010, Monthly Notices of the Royal Astronomical Society, 406, 612, (c)
  Journal compilation {\copyright} 2010 RAS

\bibitem[{Bolton {et~al.}(2009)Bolton, Oh, \& Furlanetto}]{Bolton:2009p7229}
Bolton J.~S., Oh S.~P., Furlanetto S.~R., 2009, Monthly Notices of the Royal
  Astronomical Society, 395, 736

\bibitem[{Bolton {et~al.}(2008)Bolton, Viel, Kim, Haehnelt, \&
  Carswell}]{Bolton:2008p5404}
Bolton J.~S., Viel M., Kim T.-S., Haehnelt M.~G., Carswell R.~F., 2008, Monthly
  Notices of the Royal Astronomical Society, 386, 1131

\bibitem[{Cen {et~al.}(1994)Cen, Miralda-Escud{\'e}, Ostriker, \&
  Rauch}]{Cen:1994p7419}
Cen R., Miralda-Escud{\'e} J., Ostriker J.~P., Rauch M., 1994, Astrophysical
  Journal, 437, L9

\bibitem[{Choudhury {et~al.}(2001)Choudhury, Srianand, \&
  Padmanabhan}]{Choudhury:2001p4404}
Choudhury T.~R., Srianand R., Padmanabhan T., 2001, The Astrophysical Journal,
  559, 29

\bibitem[{Cole {et~al.}(2005)}]{Cole:2005p5206}
Cole S., {et~al.}, 2005, Monthly Notices of the Royal Astronomical Society,
  362, 505

\bibitem[{Coles \& Jones(1991)}]{Coles:1991p4666}
Coles P., Jones B., 1991, Royal Astronomical Society, 248, 1

\bibitem[{Croft {et~al.}(2002)Croft, Weinberg, Bolte, Burles, Hernquist, Katz,
  Kirkman, \& Tytler}]{Croft:2002p6751}
Croft R. A.~C., Weinberg D.~H., Bolte M., Burles S., Hernquist L., Katz N.,
  Kirkman D., Tytler D., 2002, The Astrophysical Journal, 581, 20

\bibitem[{Desjacques \& Nusser(2005)}]{Desjacques:2005p3507}
Desjacques V., Nusser A., 2005, Monthly Notices of the Royal Astronomical
  Society, 361, 1257

\bibitem[{Desjacques {et~al.}(2007)Desjacques, Nusser, \&
  Sheth}]{Desjacques:2007p7310}
Desjacques V., Nusser A., Sheth R.~K., 2007, Monthly Notices of the Royal
  Astronomical Society, 374, 206

\bibitem[{Eisenstein \& Hu(1998)}]{Eisenstein:1998p3096}
Eisenstein D.~J., Hu W., 1998, Astrophysical Journal v.496, 496, 605, (c) 1998:
  The American Astronomical Society

\bibitem[{Eisenstein {et~al.}(2005)}]{Eisenstein:2005p5087}
Eisenstein D.~J., {et~al.}, 2005, The Astrophysical Journal, 633, 560

\bibitem[{Eisenstein {et~al.}(2011)}]{Eisenstein:2011p7600}
---, 2011, eprint arXiv, 1101, 1529, submitted to the Astronomical Journal

\bibitem[{Fluke {et~al.}(2010)Fluke, Barnes, Barsdell, \&
  Hassan}]{Fluke:2010p6029}
Fluke C.~J., Barnes D.~G., Barsdell B.~R., Hassan A.~H., 2010, arXiv,
  astro-ph.IM

\bibitem[{Gnedin \& Hui(1996)}]{Gnedin:1996p295}
Gnedin N.~Y., Hui L., 1996, Astrophysical Journal Letters v.472, 472, L73, (c)
  1996: The American Astronomical Society

\bibitem[{Gnedin \& Hui(1998)}]{Gnedin:1998p83}
---, 1998, Monthly Notices of the Royal Astronomical Society, 296, 44

\bibitem[{Hernquist {et~al.}(1996)Hernquist, Katz, Weinberg, \&
  Miralda-Escud{\'e}}]{Hernquist:1996p4477}
Hernquist L., Katz N., Weinberg D.~H., Miralda-Escud{\'e} J., 1996,
  Astrophysical Journal Letters v.457, 457, L51

\bibitem[{Hui {et~al.}(1997)Hui, Gnedin, \& Zhang}]{Hui:1997p4783}
Hui L., Gnedin N.~Y., Zhang Y., 1997, Astrophysical Journal v.486, 486, 599

\bibitem[{H{\"u}tsi(2006)}]{Hutsi:2006p6079}
H{\"u}tsi G., 2006, Astronomy and Astrophysics, 449, 891

\bibitem[{Kim {et~al.}(2007)Kim, Bolton, Viel, Haehnelt, \&
  Carswell}]{Kim:2007p3619}
Kim T.-S., Bolton J.~S., Viel M., Haehnelt M.~G., Carswell R.~F., 2007, Monthly
  Notices of the Royal Astronomical Society, 382, 1657

\bibitem[{Kim {et~al.}(2004)Kim, Viel, Haehnelt, Carswell, \&
  Cristiani}]{Kim:2004p7039}
Kim T.-S., Viel M., Haehnelt M.~G., Carswell R.~F., Cristiani S., 2004, Monthly
  Notices of the Royal Astronomical Society, 347, 355

\bibitem[{Kitaura {et~al.}(2010)Kitaura, Gallerani, \&
  Ferrara}]{Kitaura:2010p7306}
Kitaura F.-S., Gallerani S., Ferrara A., 2010, arXiv, astro-ph.CO

\bibitem[{Lidz {et~al.}(2010)Lidz, Faucher-Gigu{\`e}re, Dall'Aglio, McQuinn,
  Fechner, Zaldarriaga, Hernquist, \& Dutta}]{Lidz:2010p7574}
Lidz A., Faucher-Gigu{\`e}re C.-A., Dall'Aglio A., McQuinn M., Fechner C.,
  Zaldarriaga M., Hernquist L., Dutta S., 2010, The Astrophysical Journal, 718,
  199

\bibitem[{Matarrese \& Mohayaee(2002)}]{Matarrese:2002p3671}
Matarrese S., Mohayaee R., 2002, Monthly Notices of the Royal Astronomical
  Society, 329, 37

\bibitem[{McDonald \& Eisenstein(2007)}]{McDonald:2007p6752}
McDonald P., Eisenstein D.~J., 2007, Physical Review D, 76, 63009

\bibitem[{McDonald {et~al.}(2000)McDonald, Miralda-Escud{\'e}, Rauch, Sargent,
  Barlow, Cen, \& Ostriker}]{McDonald:2000p388}
McDonald P., Miralda-Escud{\'e} J., Rauch M., Sargent W. L.~W., Barlow T.~A.,
  Cen R., Ostriker J.~P., 2000, The Astrophysical Journal, 543, 1

\bibitem[{McDonald {et~al.}(2005)}]{McDonald:2005p408}
McDonald P., {et~al.}, 2005, The Astrophysical Journal, 635, 761, (c) 2005: The
  American Astronomical Society

\bibitem[{McQuinn {et~al.}(2009)McQuinn, Lidz, Zaldarriaga, Hernquist, Hopkins,
  Dutta, \& Faucher-Gigu{\`e}re}]{McQuinn:2009p7230}
McQuinn M., Lidz A., Zaldarriaga M., Hernquist L., Hopkins P.~F., Dutta S.,
  Faucher-Gigu{\`e}re C.-A., 2009, The Astrophysical Journal, 694, 842

\bibitem[{McQuinn \& White(2011)}]{McQuinn:2011p10709}
McQuinn M., White M., 2011, Monthly Notices of the Royal Astronomical Society,
  727

\bibitem[{Meiksin(2009)}]{Meiksin:2009p6791}
Meiksin A.~A., 2009, Reviews of Modern Physics, 81, 1405

\bibitem[{Norman {et~al.}(2009)Norman, Paschos, \& Harkness}]{Norman:2009p5576}
Norman M.~L., Paschos P., Harkness R., 2009, Journal of Physics: Conference
  Series, 180, 2021

\bibitem[{Padmanabhan {et~al.}(2007)}]{Padmanabhan:2007p6159}
Padmanabhan N., {et~al.}, 2007, Monthly Notices of the Royal Astronomical
  Society, 378, 852

\bibitem[{Percival {et~al.}(2007)Percival, Cole, Eisenstein, Nichol, Peacock,
  Pope, \& Szalay}]{Percival:2007p5923}
Percival W.~J., Cole S., Eisenstein D.~J., Nichol R.~C., Peacock J.~A., Pope
  A.~C., Szalay A.~S., 2007, Monthly Notices of the Royal Astronomical Society,
  381, 1053

\bibitem[{Percival {et~al.}(2010)}]{Percival:2010p6176}
Percival W.~J., {et~al.}, 2010, Monthly Notices of the Royal Astronomical
  Society, 401, 2148

\bibitem[{Rauch(1998)}]{Rauch:1998p4563}
Rauch M., 1998, Annual Review of Astronomy and Astrophysics, 36, 267

\bibitem[{Reisenegger \& Miralda-Escude(1995)}]{Reisenegger:1995p7082}
Reisenegger A., Miralda-Escude J., 1995, Astrophysical Journal v.449, 449, 476

\bibitem[{Richards {et~al.}(2006)}]{Richards:2006p8099}
Richards G.~T., {et~al.}, 2006, The Astronomical Journal, 131, 2766

\bibitem[{Schaye {et~al.}(2000)Schaye, Theuns, Rauch, Efstathiou, \&
  Sargent}]{Schaye:2000p7569}
Schaye J., Theuns T., Rauch M., Efstathiou G., Sargent W. L.~W., 2000, Monthly
  Notices of the Royal Astronomical Society, 318, 817

\bibitem[{Schlegel {et~al.}(2009)Schlegel, White, \&
  Eisenstein}]{Schlegel:2009p6826}
Schlegel D., White M., Eisenstein D., 2009, Astro2010: The Astronomy and
  Astrophysics Decadal Survey, 2010, 314

\bibitem[{Seo \& Eisenstein(2007)}]{Seo:2007p7988}
Seo H.-J., Eisenstein D.~J., 2007, The Astrophysical Journal, 665, 14

\bibitem[{Slosar {et~al.}(2009)Slosar, Ho, White, \& Louis}]{Slosar:2009p5710}
Slosar A., Ho S., White M., Louis T., 2009, Journal of Cosmology and
  Astroparticle Physics, 10, 019

\bibitem[{Slosar {et~al.}(2011)}]{Slosar:2011p8013}
Slosar A., {et~al.}, 2011, arXiv, astro-ph.CO

\bibitem[{Springel(2005)}]{Springel:2005p7117}
Springel V., 2005, Monthly Notices of the Royal Astronomical Society, 364, 1105

\bibitem[{Theuns {et~al.}(1998)Theuns, Leonard, Efstathiou, Pearce, \&
  Thomas}]{Theuns:1998p7245}
Theuns T., Leonard A., Efstathiou G., Pearce F.~R., Thomas P.~A., 1998, Monthly
  Notices of the Royal Astronomical Society, 301, 478

\bibitem[{Valageas {et~al.}(2002)Valageas, Schaeffer, \&
  Silk}]{Valageas:2002p7220}
Valageas P., Schaeffer R., Silk J., 2002, Astronomy and Astrophysics, 388, 741

\bibitem[{Viel {et~al.}(2009)Viel, Bolton, \& Haehnelt}]{Viel:2009p7711}
Viel M., Bolton J.~S., Haehnelt M.~G., 2009, Monthly Notices of the Royal
  Astronomical Society: Letters, 399, L39

\bibitem[{Viel {et~al.}(2004)Viel, Haehnelt, \& Springel}]{Viel:2004p7045}
Viel M., Haehnelt M.~G., Springel V., 2004, Monthly Notices of the Royal
  Astronomical Society, 354, 684

\bibitem[{Viel {et~al.}(2002{\natexlab{a}})Viel, Matarrese, Mo, Haehnelt, \&
  Theuns}]{Viel:2002p35}
Viel M., Matarrese S., Mo H.~J., Haehnelt M.~G., Theuns T., 2002{\natexlab{a}},
  Monthly Notices of the Royal Astronomical Society, 329, 848

\bibitem[{Viel {et~al.}(2002{\natexlab{b}})Viel, Matarrese, Mo, Theuns, \&
  Haehnelt}]{Viel:2002p4288}
Viel M., Matarrese S., Mo H.~J., Theuns T., Haehnelt M.~G., 2002{\natexlab{b}},
  MNRAS, 336, 685

\bibitem[{White {et~al.}(2010)White, Pope, Carlson, Heitmann, Habib, Fasel,
  Daniel, \& Lukic}]{White:2010p5647}
White M., Pope A., Carlson J., Heitmann K., Habib S., Fasel P., Daniel D.,
  Lukic Z., 2010, The Astrophysical Journal, 713, 383

\bibitem[{Zel'Dovich(1970)}]{ZelDovich:1970p7102}
Zel'Dovich Y.~B., 1970, Astron. Astrophys., 5, 84, a{\&}AA ID. AAA003.061.016

\bibitem[{{Zhan} {et~al.}(2005){Zhan}, {Knox}, {Tyson}, \&
  {Margoniner}}]{2005AAS...207.2605Z}
{Zhan} H., {Knox} L., {Tyson} J.~A., {Margoniner} V., 2005, in Bulletin of the
  American Astronomical Society, Vol.~37, American Astronomical Society Meeting
  Abstracts, pp. 1202--+

\end{thebibliography}

\end{document}